\documentstyle[12pt,psfig]{article}
\def\beq{\begin{equation}}
\def\eeq{\end{equation}}
\def\bea{\begin{eqnarray}}
\def\eea{\end{eqnarray}}
\def\bq{\begin{quote}}
\def\eq{\end{quote}}
\def\nn{\nonumber}

\def\gappeq{\mathrel{\rlap
{\raise.5ex\hbox{$>$}}
{\lower.5ex\hbox{$\sim$}}}}

\def\lappeq{\mathrel{\rlap{\raise.5ex\hbox{$<$}}
{\lower.5ex\hbox{$\sim$}}}}
\def\simlt{\stackrel{<}{{}_\sim}}
\def\simgt{\stackrel{>}{{}_\sim}}

\begin{document}
\pagestyle{empty}
\begin{flushright}
{CERN-TH/98-119\\
 IFT-98/22\\
 SCIPP-98/30\\
 SFB-375/305, TUM-HEP-324/98\\
hep-ph/9808275}
\end{flushright}
\vspace*{5mm}
\begin{center}
{\bf
HAGGLING OVER THE FINE-TUNING PRICE OF LEP} \\
\vspace*{1cm} 
Piotr H. Chankowski$^{a,b)}$, John Ellis$^{c)}$, 
Marek Olechowski$^{d,b)}$ and Stefan Pokorski$^{c,b)}$
\\
\vspace*{2cm}  
{\bf ABSTRACT} \\
\end{center}
\vspace*{5mm}
\noindent
We amplify previous discussions of the fine-tuning price to be paid by 
supersymmetric models in the light of LEP data, 
especially the lower bound on the Higgs boson mass,  studying in particular 
its power of discrimination between different parameter regions and different
theoretical assumptions.  The analysis is performed using the full
one-loop effective potential. The whole range of $\tan\beta$ is discussed, 
including  large values.  In the minimal supergravity model 
with universal gaugino 
and scalar masses, a small fine-tuning price is possible only for intermediate 
values of $\tan\beta$. However, the fine-tuning price in this region is
significantly higher if we require $b-\tau$ Yukawa-coupling unification.
On the other hand, price reductions are obtained if some theoretical 
relation between MSSM parameters is assumed, in particular between $\mu_0$, 
$M_{1/2}$ and $A_0$.
Significant price reductions are obtained for large 
$\tan\beta$ if non-universal soft Higgs mass parameters are allowed.
Nevertheless, in all these cases, the requirement of 
small fine tuning remains an important constraint on the superpartner spectrum.
We also study input relations between MSSM parameters suggested in some 
interpretations of string theory: the price may depend significantly on these 
inputs, potentially providing guidance for building string models. However, 
in the available models the fine-tuning price may not be reduced significantly.

\vspace*{1.5cm} 
\noindent

\rule[.1in]{16.5cm}{.002in}

\noindent
$^{a)}$ SCIPP, University of California, Santa Cruz.\\
$^{b)}$ Institute of Theoretical Physics, Warsaw University.\\
$^{c)}$ Theory Division, CERN, CH 1211 Geneva 23, Switzerland.\\
$^{d)}$ Institute f\"ur Theoretische Physik, Technische 
Universit\"at M\"unchen.
\vspace*{0.5cm}

\vfill\eject

\setcounter{page}{1}
\pagestyle{plain}

\section{Introduction}

To paraphrase Saint Augustine: "May Nature reveal supersymmetry, but 
not yet." This seems to be the message from LEP and other accelerator 
experiments, so far. There are tantalizing pieces of circumstantial 
evidence for supersymmetry at accessible energies, including the measured 
magnitudes of the gauge coupling strengths~\cite{SGUT,CHPLPO} and the 
increasing indication that the Higgs boson may be relatively 
light~\cite{EWHIGGS,LEPEWWG}. On the 
other hand, direct searches at LEP and elsewhere have so far come up 
empty-handed. In the case of LEP~2, the physics reach for many sparticles has 
almost been saturated, though the direct Higgs search still has excellent
prospects.

In this context, it is natural to wonder whether the continuing
absence of sparticles should disconcert advocates of the Minimal 
Supersymetric Extension of the Standard Model (MSSM).
After all, the only theoretical motivation for the appearance of
sparticles at accessible energies is in order to alleviate the
fine tuning required to maintain the electroweak
hierarchy~\cite{hierarchy}, and
sparticles become less effective in this task the heavier their
masses. Since the problem of fine-tuning is a subjective one,
it is not possible to provide a concise mathematical criterion
for deciding whether enough is enough, already. 
Moreover, the fine tuning can be discussed only in concrete models for
the soft supersymmetry breaking terms, and any conclusion refers to the
particular model under consideration. The fine-tuning price may also
depend on other, optional, theoretical assumptions.

The idea which we prefer to promote here is that, along with the overall
increase in the fine-tuning price imposed by the data, any sensible
objective measure of the amount of fine tuning becomes an interesting
criterion for at least comparing the relative naturalness of various 
theoretical models and constraints, and -- within a given framework -- of 
different parameter regions.

We have recently shown that the latest LEP and other data which  constrain 
the MSSM parameters significantly increased 
the requisite amount of fine tuning~\cite{CEP} compared with pre-LEP days. We used one particular 
measure~\cite{FINETUNE,BAGI}, namely $\Delta_0 \equiv {\rm max}_i
|a_i/M_Z^2(\partial M_Z^2/\partial a_i)|$, where the $a_i$ are the 
input parameters of the MSSM (for other measures of fine tuning, see 
\cite{OTHERS}).
Our tree-level analysis clearly demonstrated several qualitative
trends but, as an obvious improvement,  one should use the best 
available theoretical tools to evaluate the fine tuning, including in
particular the full one-loop effective potential of the MSSM~
\cite{OLPO,BAST}. 
Secondly, one should update the analysis with the most recent experimental 
information, in particular on the mass of the Higgs boson~\cite{LEPHiggs}
and the new result for $BR(b\rightarrow s\gamma)$ \cite{CLEOnew}.

With the above improvements in hand, in this paper we address 
anew the question 
of the necessary amount of fine tuning,  with a particular 
view to the power of the fine-tuning price to discriminate between
different parameter regions and different theoretical assumptions. 

In Section 2 we recall our measure of fine tuning and discuss its various 
qualitative aspects in the supergravity-mediated scenario with universal
gaugino and scalar masses (the minimal supergravity model). 
The particular role of the Higgs boson mass
is elucidated. In Section 3 we present our full one-loop results for small 
and intermediate $\tan\beta$ in this model. In the first place, 
we confirm previous 
findings \cite{OLPO,BAST} that including the full one-loop effective 
potential reduces the apparent amount of fine-tuning by about 30\% at
moderate $\tan\beta\sim10$, and by much  larger factors for both small and large 
$\tan\beta$. On the other hand, the latest experimental lower limit 
$M_h>90$~GeV for low $\tan\beta$ increases the price again, so that the 
fine-tuning price we find for low $\tan\beta$ is not very different from 
that in~\cite{CEP}. In the minimal supergravity model, for intermediate 
$\tan\beta$: $3<\tan\beta<15$, there 
still exist domains of the parameter space with moderate, ${\cal O}(10\%)$, 
fine-tuning. This result is obtained after including all available experimental
constraints, including in particular $b\rightarrow s \gamma$, but with no 
constraint on the Yukawa sector. 

In Section 4 we address the question of bottom-quark/tau Yukawa-coupling 
($b-\tau$) unification~\cite{BUEL} in the minimal model for small and
intermediate $\tan\beta \lappeq 30$. We show that inclusion of one-loop
 corrections to the bottom-quark mass substantially enlarges the
$\tan\beta$
region where $b-\tau$ unification is possible, albeit at the expense of 
a higher fine-tuning price. Furthermore, the interplay of the 
constraints from $b-\tau$ 
unification and $b\rightarrow s \gamma$ decay and the dependence on
$\tan\beta$ is understood. 
The minimal model with both $b-\tau$ unification and the $b \rightarrow s
\gamma$ constraint imposed has no regions of small fine
tuning. However, it is stressed that $b\rightarrow s \gamma$ decay
is an optional constraint, which can be relaxed if we admit some 
departure from the minimal model, e.g., some flavour structure in the up-squark 
mass matrices. Given such a generalization,
regions with low fine tuning exist with or without $b-\tau$ unification.

In Section 5 we  emphasize the dependence of the fine-tuning price
on the choice of the set of independent soft mass parameters in a given model.
In particular, we find that in the minimal supergravity model $\Delta_0$
may be 
significantly reduced if the parameters $\mu_0$ and $M_{1/2}$ or $A_0$ 
(depending 
on the value of $\tan\beta$) are considered as linearly dependent on
each other. In some stringy models these parameters are indeed not independent,
although the correlation may not be linear. As is  briefly discussed in 
Section 6, we find that in one class of such models $\Delta_0$ may be
minimized only in unphysical regions of the parameter space 
corresponding to small sparticle masses and/or the absence of electroweak 
symmetry breaking. Within the physical region of parameters in the models 
studied, the fine-tuning price may not be reduced significantly. 

In Section 7  we discuss the case of large $\tan\beta > 30$. 
Our main conclusion is that it remains attractive (with small 
fine-tuning) for non-universal Higgs boson mass parameters at the GUT scale. 
Section 8 contains our conclusions.

\section{Measure of Fine Tuning and Tree-Level Discussion}

We first specify more precisely the fine-tuning criterion we use.
Following \cite{FINETUNE,BAGI}, we consider the logarithmic sensitivities of 
$M_Z$ with respect to variations in input parameters $a_i$:
\begin{equation}
\Delta_{a_i}={a_i\over M_Z}{\partial M_Z\over\partial a_i}
\label{eqn:derivatives}
\end{equation}
Note that here we take derivatives of $M_Z$ and not, as in~\cite{CEP},
of 
$M^2_Z$:  hence our $\Delta_{a_i}$  are smaller by factors 2, other things being
equal. We then define
\begin{equation}
\Delta_0\equiv {\rm max} |\Delta_{a_i}|
\label{banana}
\end{equation}
It is clear that the fine tuning can be discussed only in concrete models
for the soft supersymmetry breaking terms, and with a specified scale for their
generation. Calculation of the derivatives (\ref{eqn:derivatives}) requires 
minimization of the effective scalar potential written in terms of the 
$a_i$. In the first approximation one may use (as we did in our previous 
paper) the tree-level form of the potential, but it is known \cite{OLPO} and 
has been strongly re-emphasized recently \cite{BAST} that reliable 
quantitative analysis 
requires use of the full one-loop effective potential. This is
particularly 
important for low and large values of $\tan\beta$, where one-loop corrections 
to the tree-level potential are decisive for electroweak symmetry 
breaking. In this paper we follow the one-loop approach of \cite{OLPO}.

It is, nevertheless, useful to discuss first certain qualitative features
of our 
analysis starting with the tree-level potential:
\begin{eqnarray}
V&=&m^2_1|H_1|^2+m_2^2|H_2|^2 + B\mu
(\epsilon H_1H_2+\epsilon\bar H_1\bar H_2)\nonumber\\
 &+&{1\over8}(g^2+g^{\prime2})(|H_1|^2-|H_2|^2)^2 
 + {g^2\over2}|\bar H_1H_2|^2
\end{eqnarray}
The derivatives (\ref{eqn:derivatives}) then read 
\bea
\Delta_{a_i}={2\over(t^2-1)^2}\sum_j&&\!\!\!\!\!\!\!\left\{
\left(c_B^ic_\mu^j+c_B^jc_\mu^i\right)
t(t^2+1)\left[{a_ia_j\over M_Z^2}+{a_ia_j\over M_A^2}\right]\right.\nn\\
&&\left.\!\!\!-c_1^{ij}
\left[(t^2+1){a_ia_j\over M_Z^2}+2t^2{a_ia_j\over M_A^2}\right]
-c_2^{ij}~t^2
\left[(t^2+1){a_ia_j\over M_Z^2}+2{a_ia_j\over M_A^2}\right]\right\}
\label{eq:Delta_ai}
\eea
where $t\equiv\tan\beta$ and the coefficients $c_{1,2}^{ij}$, $c_B^i$ and 
$c^i_\mu$ are defined by
\bea
m_{1,2}^2 \left(M_Z^2\right) =
&&\!\!\!\!\!\!\!\!
\sum_{ij}c_{1,2}^{ij}a_ia_j\phantom{aaaa}c_{1,2}^{ij}=c_{1,2}^{ji}\nn\\
B\left(M_Z^2\right)=
&&\!\!\!\!\!\!\!\!
\sum_i c_B^i a_i
\\
\mu \left(M_Z^2\right)
=
&&\!\!\!\!\!\!\!\!
\sum_i c_\mu^i a_i
\nn
\eea
The numerical values of the coefficients $c_{1,2}^{ij}$, $c_B^i$ and 
$c^i_\mu$ can be found by solving the renormalization-group (RG) equations~\cite{CAOLPOWA1,CACHOLPOWA} 
for the running from some initial scale down to $M_Z$. 

The most popular model, with some phenomenological backing, is the MSSM with 
universal gaugino and scalar masses $(M_{1/2}, m_0)$ at the input supergravity
scale (or GUT scale), and a universal trilinear (bilinear) supersymmetry 
breaking parameter $A_0(B_0)$. The model is then formulated in terms of 
these four parameters and the $\mu_0$ 
parameter. Important features of this 
model are strong correlations between soft terms: all scalar mass parameters 
are assumed to be equal. At present we do not have any convincing theory of 
soft terms and might equally well contemplate the possibility of other
patterns for 
them, for instance of non-universal Higgs-boson masses and/or different
sets of independent 
parameters. The amount of fine tuning  depends on this choice, as discussed in
Section~5.  However, in this and the following Section we discuss the universal case
with five independent parameters (the minimal supergravity model).
 
Several qualitative effects 
in the fine tuning of soft terms can be seen already from (\ref{eq:Delta_ai}). 
In particular, we can discuss the typical magnitude of the
derivatives taken with respect to the five parameters of the minimal
supergravity model, with the scale of the generation of soft terms taken 
to be $M_{GUT}=2\times10^{16}$ GeV.
As an example, we consider the region of small $\tan\beta$, not far from 
the quasi-infrared fixed-point solution for the top-quark Yukawa coupling,
in which the fine tuning is generically larger than for intermediate values
of $\tan\beta$.

Using analytic solutions obtained in \cite{CAOLPOWA1} 
for the coefficients $c_{1,2}^{ij}$, $c_B^i$ and $c^i_\mu$ 
as an expansion in the parameter 
\begin{eqnarray}
y\equiv Y_t/Y_t^{FP}
\end{eqnarray}
where $Y_t^{FP}$ is the fixed-point value for the top-quark Yukawa coupling
$Y_t\equiv h^2_t/4\pi$, we can calculate the derivatives in 
eq.~(\ref{eq:Delta_ai}) explicitly. In the limit $y\rightarrow1$ one gets:
\begin{eqnarray}
\Delta_{\mu_0}=-{2\over (t^2-1)^2} \left[\left(t^2+1\right)^2{\mu^2\over M^2_Z}
+ t^2\left(4 {\mu^2\over M^2_A} -1-
{M_A^2\over M^2_Z}\right)\right]\nonumber
\end{eqnarray}
\begin{eqnarray}
\Delta_{M_{1/2}}\approx{t^2+1\over(t^2-1)^2}\left[\left(7t^2-1\right)
{M_{1/2}^2\over M^2_Z}
+ t\left({\mu M_{1/2}\over M^2_Z} + {\mu M_{1/2}\over M^2_A}\right)
+{12t^2\over t^2+1}{M_{1/2}^2\over M^2_A}\right]\nonumber
\end{eqnarray}
\begin{eqnarray}
\Delta_{A_0}\approx-{t(t^2+1)\over(t^2-1)^2}\left[{\mu A_0\over M^2_Z}
+ {\mu A_0\over M^2_A}\right]
\label{eqn:univdeltas}
\end{eqnarray}
\begin{eqnarray}
\Delta_{B_0}\approx {2t(t^2+1)\over (t^2-1)^2}\left({\mu B_0\over M^2_Z}
+ {\mu B_0\over M^2_A}\right)\nonumber
\end{eqnarray}
\begin{eqnarray}
\Delta_{m_0} \approx {1\over(t^2-1)^2}\left[(t^2+1)(t^2-2)
{m_0^2\over M^2_Z} -2t^2 {m_0^2\over M^2_A}\right]\nonumber
\end{eqnarray}
where we may consider the region $t\equiv\tan\beta\sim(1.5-2)$ to be
consistent with our approximation $y\approx1$ - 
qualitatively similar conclusions can be drawn for 
other values of $\tan\beta$, as long as the bottom quark Yukawa coupling is
much  smaller than $Y_t$.
The parameter $\mu$ at the
scale $M_Z$ is related to its initial value $\mu_0$ by the equation
$\mu^2\approx2\mu^2_0(1-y)^{1/2}$. We note that 
the largest derivatives are $\Delta_{\mu_0}$ and 
$\Delta_{M_{1/2}}$, and they are of opposite signs.
For instance, for all parameters of order
$M_Z$ (a situation already strongly excluded by the present experimental 
constraints)  both are already greater than $\sim10$ for
$\tan\beta\sim 1.5$. They increase quadratically with the values of the 
parameters, and this is the reason why large fine tuning is found  for 
low $\tan\beta$
within the experimentally-acceptable  parameter range. The derivative
$\Delta_{A_0}$ is also sizeable and may also play an important role, since large
negative $A_0$ may be necessary \cite{CACHPOWA} to satisfy 
the present Higgs-boson mass
limit \cite{LEPHiggs}. The derivative $\Delta_{m_0}$ is typically
of little importance. For a given parameter set, the necessary fine tuning
is determined by the maximal derivative
(\ref{banana}). Any such set should be chosen consistent
with the present experimental data and the constraints (correlations) imposed 
by  proper electroweak symmetry breaking (see for 
instance \cite{OLPO,KANE}).

Among the experimental constraints, a special role is played by the 
Higgs-boson mass limits. This effect can be isolated by imposing all the available 
experimental constraints except the lower limit on $M_h$. For a chosen
value of $\tan\beta$,  we get then some minimal value of $\Delta_0$, and
the
corresponding parameter set determines the mass of the Higgs boson 
\begin{eqnarray}
M^2_h=M^2_Z\cos^22\beta+{3\alpha\over4\pi s^2_W}{m^4_t\over M^2_W}
\left[\log\left({M^2_{\tilde t_2}M^2_{\tilde t_1}\over m^4_t}\right)
+\left({M^2_{\tilde t_2}-M^2_{\tilde t_1}\over4m^2_t}\sin^22\theta_{\tilde t}
\right)^2\right.\nonumber
\end{eqnarray}
\begin{eqnarray}
\times\left. f(M^2_{\tilde t_2},M^2_{\tilde t_1})
+ {M^2_{\tilde t_2}-M^2_{\tilde t_1}\over2m^2_t}\sin^22\theta_{\tilde t}
\log\left({M^2_{\tilde t_2}\over M^2_{\tilde t_1}}\right)\right]
\label{eqn:mhha}
\end{eqnarray}
where $f(x,y)\equiv 2 - (x + y)/(x - y)\log(x/y)$. We display in
(\ref{eqn:mhha}) only the 
one-loop formula valid for $M_A\simgt3M_Z$ \cite{HE},
which is a good approximation to the two-loop RG-improved prescription 
\cite{CAESQUWA} used in our numerical studies. Due to the logarithmic 
dependence of $M_h^2$ on the physical stop masses squared (which in turn
are functions of the initial parameters) any departure from
the ``best'' value of $M_h$, for fixed $\tan\beta$ and in the range allowed
by the other constraints, transmits itself into an approximately exponential
rise of $\Delta_0$. This happens for $M_h$ changing in both directions, 
towards values both smaller and larger than the ``best'' value. 
Thus, for a given $\tan\beta$, the Higgs boson mass is a crucial probe of 
fine tuning. We also recall \cite{CAESQUWA} that the Higgs boson mass 
is maximal for large $|\tilde A_t|\equiv |A_t-\mu\cot\beta|$. Since 
$A_t$ is given by \cite{CAOLPOWA1}
\begin{eqnarray}
A_t\approx(1-y)A_0-{\cal O}(1-2)M_{1/2}\label{eqn:at},
\end{eqnarray}
maximizing $M_h$
requires $\mu>0$ and large negative $A_0$. Hence 
the derivative $\Delta_{A_0}$ may be large in the low $\tan\beta$ region.

\section{Results for Low and Intermediate $\tan\beta$}

In this Section we discuss our full one-loop  results in the minimal 
supergravity model with five independent parameters.
It is appropriate to begin by re-emphasizing that the fine-tuning 
criterion is not a rigorous mathematical statement, but rather an intuitive 
physical preference and hence remains necessarily subjective.
(For instance, as already mentioned, our present definition differs by a 
factor 2 from the one used in 
\cite{CEP}.) Nevertheless, for a chosen  measure of  fine-tuning, one 
can  study in this model relative changes in the amount of fine tuning as 
a function of changing experimental limits and of the considered parameter
range. Here we emphasize this use of the naturalness criterion.

We first recall the experimental constraints used in this analysis. The 
data we take into account include the precision electroweak data 
reported at the Jerusalem conference \cite{LEPEWWG}, which are dominated by 
those from LEP~1. We constrain MSSM parameters by requiring that 
$\Delta\chi^2 <4$ in a global MSSM fit \cite{ALTAR,ELLIS,KANE2,HOLLIK}.
The main effect of this constraint is a lower bound on the left-handed stop: 
$M_{\tilde t_L}\simgt300-400$ GeV \cite{KANE2}. We also take into account the 
direct LEP~2 lower limits on the masses of sparticles \cite{LEPC183} and Higgs 
bosons. For the latter, we base ourselves on the recent data reported in
 \cite{LEPHiggs}. We use the limit $M_h>90$ GeV which, 
strictly
speaking, is valid for the Standard Model Higgs boson. This is approximately
valid also for the MSSM Higgs
boson for small tan$\beta$, although there still exist small windows in
the parameter space where the 
experimental limit is lower. We neglect this possibility in the present
analysis.

\begin{figure}
\psfig{figure=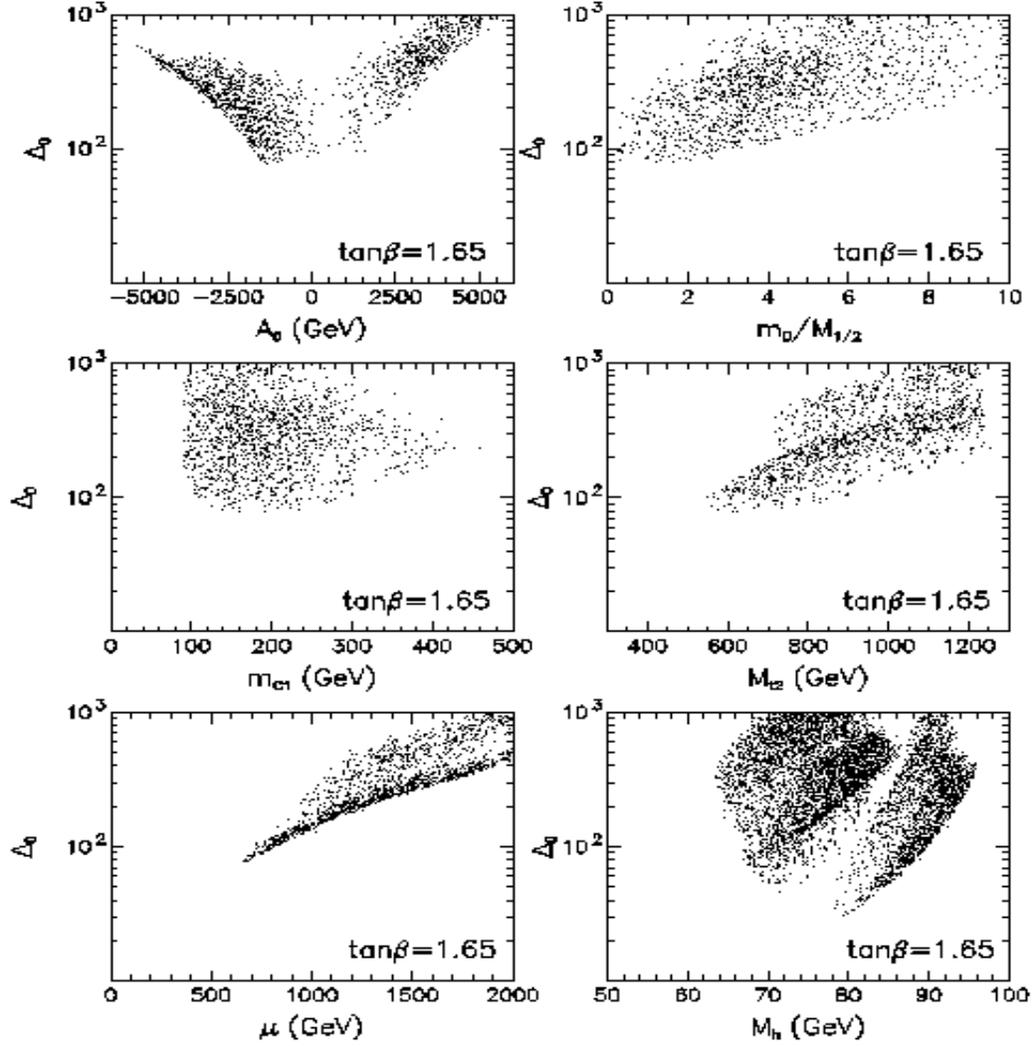,width=15.0cm,height=15.0cm}
\vspace{1.0truecm}
\caption{{\it The price of fine tuning for $\tan\beta=1.65$, as a function of 
various variables in the minimal supergravity model.  
An upper limit of 1.2 TeV on the heavier stop mass  is 
imposed in the scanning. All experimental constraints described in the text
are included. In all plots, except for $\Delta_0$ versus $M_h$, the bound
$M_h>90$ GeV is included. The mass of the lighter physical chargino and of the
heavier physical stop are denoted by $m_{C1}$ and 
$M_{\tilde t_2}$, respectively.}}
\end{figure}

The final accelerator contraint we use is the recently-measured value of 
the $b\rightarrow s\gamma$ branching ratio
$2\times10^{-4}<B(B\rightarrow X_s\gamma)<4.5\times10^{-4}$ at the 95\% C.L.
\cite{CLEOnew}. The interpretation of this measurement in the MSSM is still 
subject to some uncertainty, 
because not all the ${\cal O}(\alpha_s)$ 
corrections have yet been calculated. Resumming large QCD logaritms of the 
type $\log(M_W/m_b)$ up to next-to-leading order (NLO) accuracy has 
recently been accomplished \cite{CHMIMU}. 
These calculations are identical in the 
SM and the MSSM. The initial numerical values of the Wilson coefficients at 
the scale $\mu\approx M_W$ are, however, different in the two models. 
In our analysis we have used for 
them two-loop results available for the standard $W^\pm t$ and 
$H^\pm t$ \cite{ADYA} contributions, and only the leading-order results 
for the chargino-stop contribution \cite{BAGIbsg}. The uncertainty
due to ${\cal O}(\alpha_s/\pi)$ corrections to them has been, however,
included as in \cite{MIPORO,KANE2}. Those references also contain 
extensive discussions of the role played by the $b\rightarrow s\gamma$ 
measurement in constraining the parameter space of the MSSM. 

An important role may also be played by non-accelerator constraints, in 
particular the relic cosmological density of neutralinos $N^0$, if these
are assumed to be the lightest supersymmetric particles, and if $R$ parity
is absolutely conserved. Both of these assumptions may be disputed, and a 
complete investigation of astrophysical and cosmological constraints is beyond
the scope of this analysis (for steps in this direction, see 
\cite{ELFAGAOLSC}).

One-loop corrections to the effective potential are taken into account as 
in \cite{OLPO}, using the decoupling method of \cite{CHANK}. 
Numerical calculation of the derivatives $\Delta_{a_i}$ is 
also explained in \cite{OLPO}. Electroweak symmetry must, of course,
be broken, and this requirement imposes strong constraints on the allowed
parameter region. The main effect is that $\mu>{\rm max}(M_{1/2}, ~m_0)$. 

\begin{figure}
\psfig{figure=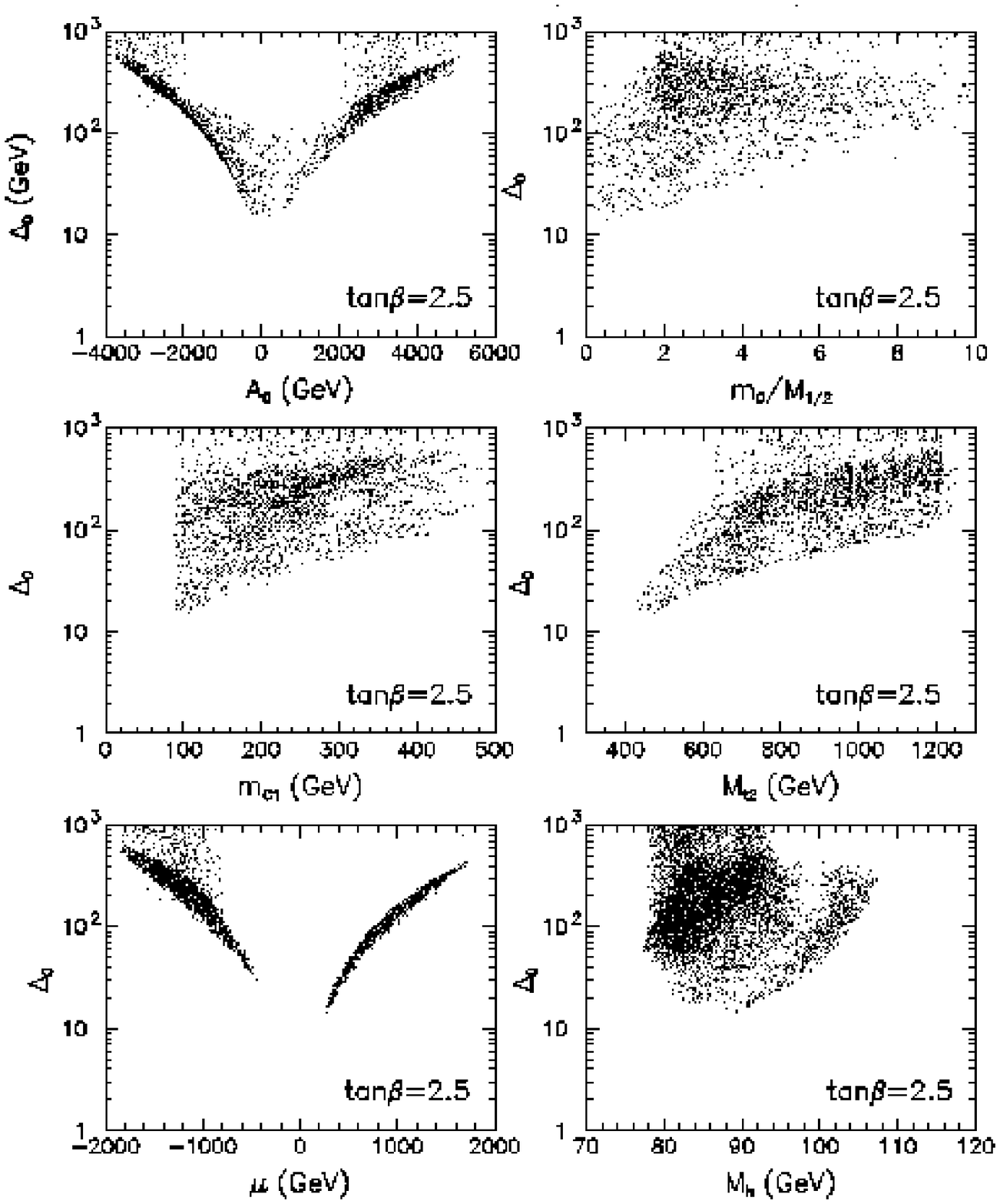,width=15.0cm,height=15.0cm}
\vspace{1.0truecm}
\caption{{\it As in Fig.~1, but for $\tan\beta=2.5$.}}
\end{figure}

\begin{figure}
\psfig{figure=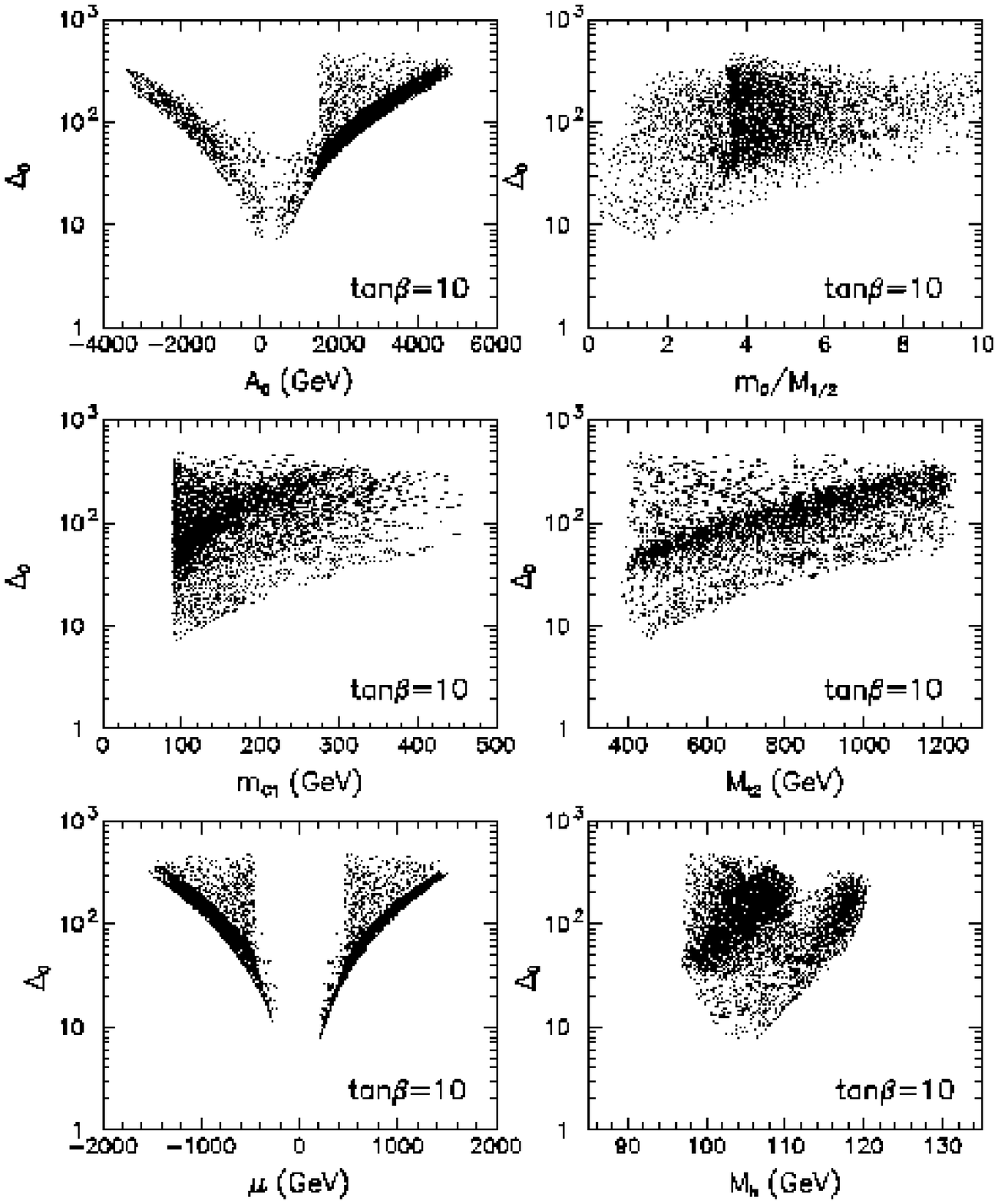,width=15.0cm,height=15.0cm}
\vspace{1.0truecm}
\caption{{\it As in Fig.~1, but for $\tan\beta=10$.}}
\end{figure}

In Figs.~1 to 3 we show $\Delta_0\equiv{\rm max}|\Delta_{a_i}|$ as a
function 
of some mass parameters and some physical masses, for $\tan\beta=1.65$, 2.5
and 10. The results are in agreement with the qualitative discussion of
Section 2 and with the results of \cite{CEP}, but the inclusion of one-loop 
corrections to the scalar potential sizeably decreases the
fine-tuning price. For the same experimental constraints (in particular
the same lower limit on $M_h$), the minimal value of $\Delta_0$ is for 
$\tan\beta=1.65$  a factor of $\sim 3-4$ smaller than in \cite{CEP}
(remember the factor 2 in the present definition of $\Delta_0$), in agreement
with previous findings \cite{OLPO,BAST}. For $\tan\beta\sim{\cal O}(10)$, 
one-loop  corrections give much smaller effects, with typically a
30\% reduction. In Figs.~1 to 3 we observe a very strong dependence of 
$\Delta_0$ on $M_h$, which has been explained in Section 2. In consequence,
the new lower limit  $M_h\simgt90$ GeV \cite{LEPHiggs} pushes, for
$\tan\beta=1.65$, the minimal value of $\Delta_0$ into the range 
$\sim{\cal O}(100)$ ($\sim{\cal O}(200)$ with the definition of $\Delta_0$ used
in \cite{CEP}). We note the increase by a factor 3 in the minimal $\Delta_0$
with the change in the lower bound on $M_h$ from 80 to 90 GeV. 
The results in all three Figures are qualitatively similar except for the 
overall decrease in the fine-tuning price with increasing $\tan\beta$.
One more difference is that the $\mu<0$ branch of solutions has disappeared
at $\tan\beta=1.65$. For so small a value of $\tan\beta$, as explained earlier,
negative $\mu$ is no longer compatible with the present bound $M_h>90$ GeV.

\begin{figure}
\psfig{figure=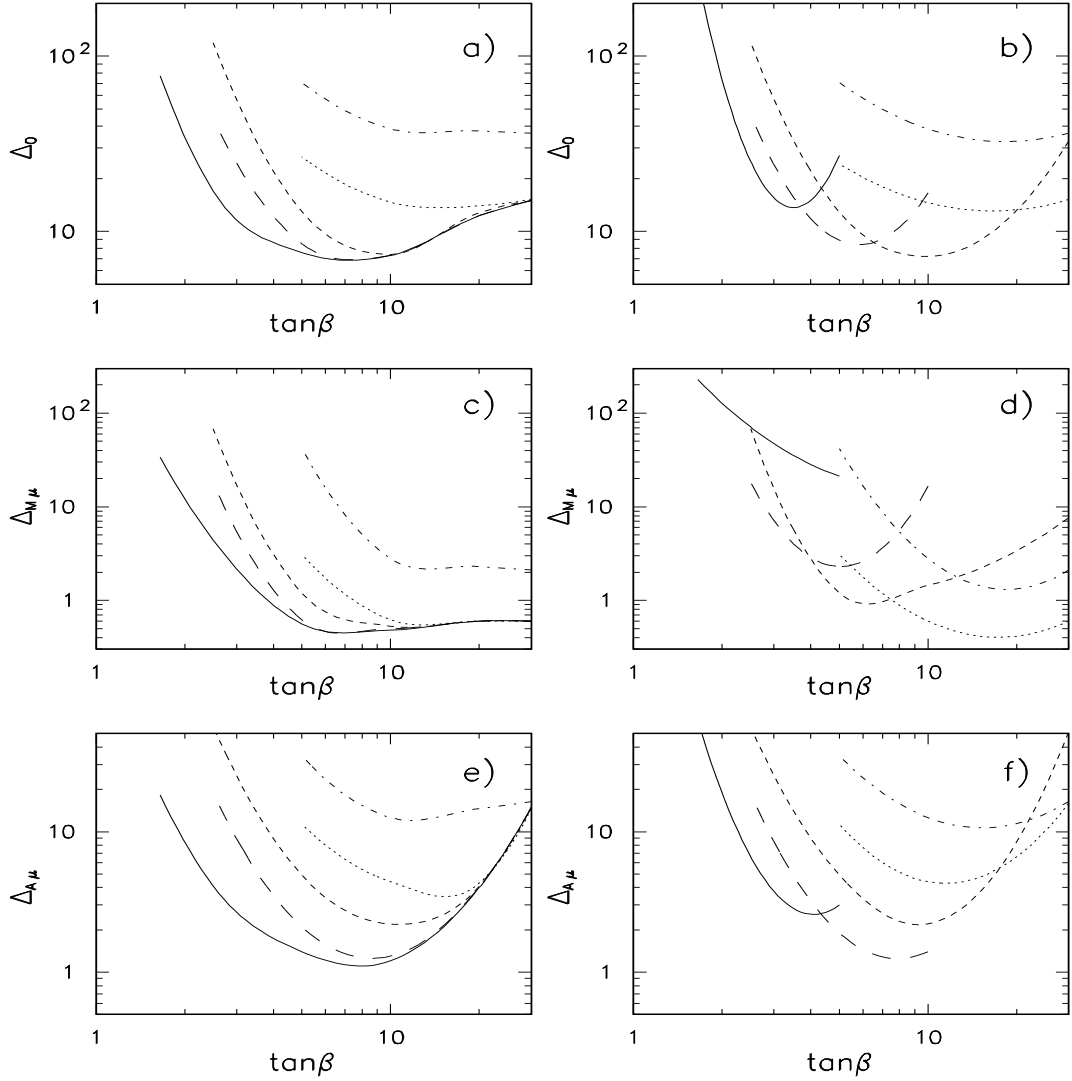,width=15.0cm,height=15.0cm}
\vspace{1.0truecm}
\caption{{\it Fine-tuning measures as functions of $\tan\beta$. 
In panels (a),(c) and (e), lower limits on the Higgs boson mass of 90 GeV 
(solid), 100 GeV (long-dashed), 105 GeV (dashed) 110 GeV (dotted) and 115 GeV 
(dot-dashed) have been assumed. In panels (b), (d) and (f), $M_h$ has been
fixed to 95 GeV (solid), 100 GeV (long-dashed), 105 GeV (dashed) 110 GeV 
(dotted) and 115 GeV (dot-dashed).  Panels (a) and (b) correspond to 
independent $M_{1/2}$, $A_0$ and $\mu$ parameters. In panels (c), (d) and 
(e), (f), linear dependences $M_{1/2}=c_{M\mu}\mu_0$ and 
$A_0=c_{A\mu}\mu_0$, respectively, have been assumed.}}
\end{figure}

The  fine tuning decreases with increasing $\tan\beta$, with values of 
$\Delta_0$ marginally reaching $\Delta_0\approx10$ for $3<\tan\beta<15$. 
This is shown in Fig.~4a, where we plot (solid line) the minimal $\Delta_0$ 
as a function of $\tan\beta$ for $M_h>90$ GeV. Also in Fig.~4a we show 
similar plots, but for hypothetical lower limits on the Higgs boson mass 
$M_h>100$, 105, 110 and 115 GeV. We notice that for $M_h>115$ GeV the 
fine-tuning is large for all values of  $\tan\beta<30$. It is also 
interesting to observe that the bulk of the parameter range shown in 
Figs.~1 to 3 gives interestingly large fine tuning, even for intermediate 
values of $\tan\beta$. Finally, given the striking dependence of $\Delta_0$ 
on $M_h$, it is interesting to see its dependence on $\tan\beta$ under the 
assumption that we know the Higgs boson mass. In Fig.~4b this dependence is 
plotted for the hypothetical values
$M_h=95$, 100, 105, 110 and 115 GeV. For values $M_h\leq105$ GeV, 
$\Delta_0$ as a function of $\tan\beta$ has a clear minimum, which moves 
towards larger values of $\tan\beta$ with increasing $M_h$.

\section{Bottom-Tau Yukawa Unification and $b\rightarrow s\gamma$ Decay}

Up to now, we have been discussing fine tuning in the minimal supergravity
model, without any constraints imposed on Yukawa couplings at the GUT scale.
One important remark is that, in such a framework, the $b\rightarrow s\gamma$
decay, although constraining for the parameter space, does not have any impact
on the necessary amount of fine tuning. The results for the minimal 
$\Delta_0$ presented in Figs.~1 to 3 and 4a does not depend at all on the
inclusion of $b\rightarrow s \gamma$ decay among our experimental
constraints for $\tan\beta\simlt 15$, and are negligibly modified for
$\tan\beta$ up to 30. This is no longer true if we impose some constraints
on the Yukawa sector, which is discussed in this Section and Section 7.

One interesting possibility is  $b-\tau$ Yukawa-coupling unification at
the GUT scale \cite{BUEL}. It is well known that exact $b-\tau$ Yukawa-coupling
unification, at the level of two-loop renormalization group equations for 
the running from the GUT scale down to $M_Z$, supplemented by three-loop QCD
running down to the scale  $M_b$ of the pole mass and finite two-loop QCD 
corrections at this scale, is possible only for very small or very 
large values of $\tan\beta$. This is due to the fact that renormalization of 
the $b$-quark mass by strong interactions is too strong, 
and has to be partly compensated by a large $t$-quark Yukawa coupling.
This result is shown in Fig.~5a. We compare there the running mass $m_b(M_Z)$
obtained by the running down from $M_{GUT}$, where we take $Y_b=Y_{\tau}$,
with the range of $m_b(M_Z)$ obtained from the pole mass $M_b=(4.8\pm 0.2)$
GeV \cite{BENEKE}, taking into account the above-mentioned low-energy
corrections. These translate the range of the pole mass: 4.6 $<M_b<$ 5.0 GeV 
into the following  range of the running mass $m_b(M_Z)$: 
$2.72<m_b(M_Z)<3.16$~GeV. To remain conservative, we use 
$\alpha_s(M_Z)=0.115(0.121)$ to obtain an upper (lower) limit on  $m_b(M_Z)$.

\begin{figure}
\psfig{figure=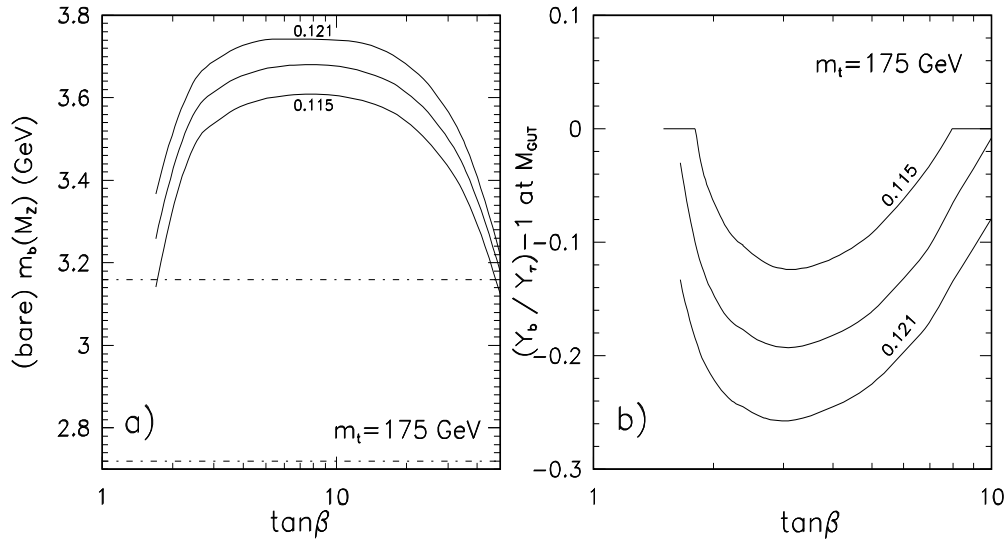,width=14.0cm,height=8.0cm}
\vspace{1.0truecm}
\caption{{\it a) The running mass $m_b(M_Z)$ obtained from strict 
$b-\tau$ Yukawa coupling unification at $M_{GUT}=2\times10^{16}$ GeV 
for different values of $\alpha_s(M_Z)$, before inclusion of one-loop
supersymmetric corrections.  b) The minimal departure from $Y_b=Y_\tau$
at $M_{GUT}$ measured by the ratio $Y_b/Y_\tau -1$, which is necessary for 
obtaining the correct $b$ mass in the minimal supergravity model with   
one-loop supersymmetric corrections included.}}
\end{figure}

It is also well known \cite{HARASA,CAOLPOWA} that, 
at least for large values of $\tan\beta$, supersymmetric
finite one-loop corrections (neglected in Fig.~5a) are very important. These
corrections are usually not considered for intermediate values of $\tan\beta$
but, as we shall demonstrate, they are also very important there and make
$b-\tau$ unification viable in much larger range of $\tan\beta$
than generally believed (see also \cite{MAPI}). However, one has then 
to pay a higher fine-tuning price!

\begin{figure}
\psfig{figure=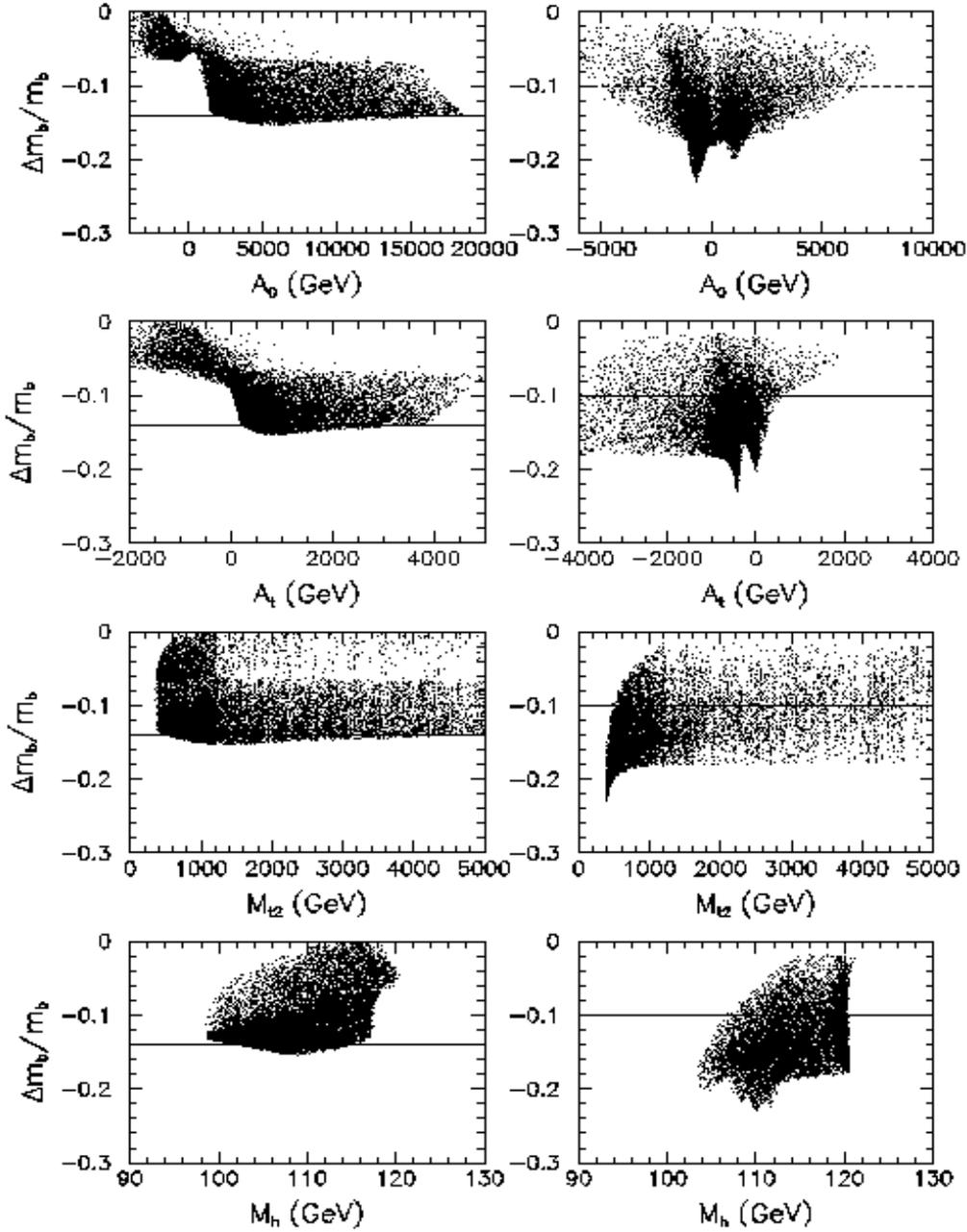,width=15.0cm,height=18.0cm}
\vspace{1.0truecm}
\caption{{\it One-loop supersymmetric corrections to the $b$-quark  
mass as functions of various parameters
for $\tan\beta=10$ (left panels) and 30 (right panels), assuming the minimal
supergravity scenario and imposing all experimental cuts except for the 
$b\rightarrow s\gamma$ constraint. Acceptable values of  $M_b$ are obtained 
for $\Delta m_b/m_b <-0.14$ for $\tan\beta=10$ and $\Delta m_b/m_b <-0.1$ for 
$\tan\beta=30$, corresponding to the regions below the solid horizontal
lines.}}
\end{figure}

One-loop diagrams with bottom squark-gluino and top squark-chargino loops 
make a contribution to the bottom-quark mass which is proportional to 
$\tan\beta$ \cite{HARASA,CAOLPOWA}. We recall that, to a good approximation,
the one-loop correction to the bottom quark mass is  given by the expression: 
\begin{eqnarray}
{\Delta m_b\over m_b} \approx 
{\tan\beta\over4\pi}\mu\left[{8\over3}\alpha_s m_{\tilde g}
I(m^2_{\tilde g},M^2_{\tilde b_1},M^2_{\tilde b_2}) + 
Y_tA_tI(\mu^2,M^2_{\tilde t_1},M^2_{\tilde t_2})\right]\label{eqn:mbcorr}
\end{eqnarray}
where 
\begin{eqnarray}
I(a,b,c)=-{ab\log(a/b)+bc\log(b/c)+ca\log(c/a)\over(a-b)(b-c)(c-a)}\nonumber
\end{eqnarray}
and the function $I(a,b,c)$ is always positive and approximately inversely 
proportional to its largest argument. This is the correction to the running 
$m_b(M_Z)$. It is clear from Fig.~5a that for $b-\tau$ unification in the  
intermediate $\tan\beta$ region we need a negative correction of order 
(15-20)\% for $3 \lappeq \tan\beta \lappeq 20$, and about a 10\% correction 
for $\tan\beta=30$. According to (\ref{eqn:mbcorr}), such corrections require 
$\mu<0$, and their dependence  on some other parameters is shown in Fig.~6.

We notice that, as expected from (\ref{eqn:mbcorr}), $b-\tau$ unification 
is easier for $\tan\beta=30$ than for $\tan\beta\approx10$. In the latter 
case it requires $A_t\simgt0$, in order to obtain an enhancement in 
(\ref{eqn:mbcorr}) or at
least to avoid any cancellation between the two terms in (\ref{eqn:mbcorr}). 
This is a strong constraint on the parameter space. Since $A_t$ is 
given by (\ref{eqn:at}), ~$b-\tau$ unification requires large positive 
$A_0$ and not too large a $M_{\tilde g}$ (i.e., $M_{1/2}$). In addition, the 
low-energy value of $A_t$ is then always relatively
small, and this explains the stronger upper bound on $M_h$ seen in Fig.~6 
(for a similar conclusion, see \cite{MAPI}). We see in Fig.~5b that, for 
$\tan\beta\simlt 10$, the possibility of exact $b-\tau$ unification evaporates 
quite quickly, with a non-unification window for $2\simlt\tan\beta\simlt8-10$, 
depending on the value of $\alpha_s$. 
However, we also see that supersymmetric one-loop corrections are large enough 
to assure unification within 10\% in almost the whole range of small and 
intermediate $\tan\beta$. 

For $\tan\beta >10$, the qualitative picture changes gradually. The overall
factor of $\tan\beta$, on the one hand, and the need for smaller corrections, 
on the other hand, lead to the situation where a partial cancellation of the
two terms in (\ref{eqn:mbcorr}) is necessary, or both corrections must 
be suppressed by sufficiently heavy squark masses.  Therefore, as seen 
in Fig.~6 , $b-\tau$ unification for $\tan\beta=30$ typically requires a 
negative value of $A_t$, and is only marginally possible for positive $A_t$, 
for heavy enough squarks. A similar but more extreme situation occurs 
for very large 
$\tan\beta$ values, which will be discussed in Section 7. It is worth 
recalling already here that the second term in (\ref{eqn:mbcorr}) 
is typically at most of order of (20-30)\% of the first term \cite{CAOLPOWA}, 
due to (\ref{eqn:at}). Thus, cancellation of the two 
terms is limited, and for very large $\tan\beta$ the contribution of 
(\ref{eqn:mbcorr}) must be anyway suppressed by requiring heavy squarks. 
This trend is visible in Fig.~7a already for $\tan\beta=30$. 
The Higgs-boson mass is not constrained by $b-\tau$ unification, since
$A_t$ can be negative and large. Finally, we observe that, for any value of
$\tan\beta$, the one-loop correction to $m_b(M_Z)$ (\ref{eqn:mbcorr}) 
remains approximately constant 
after simultaneous rescaling of $\mu$, $M_{1/2}$ and $m_0$. Since proper
electroweak breaking correlates $\mu$ with $M_{1/2}$ and $m_0$, the loop
correction to $m_b(M_Z)$ is weakly dependent on sparticle masses.

\begin{figure}
\psfig{figure=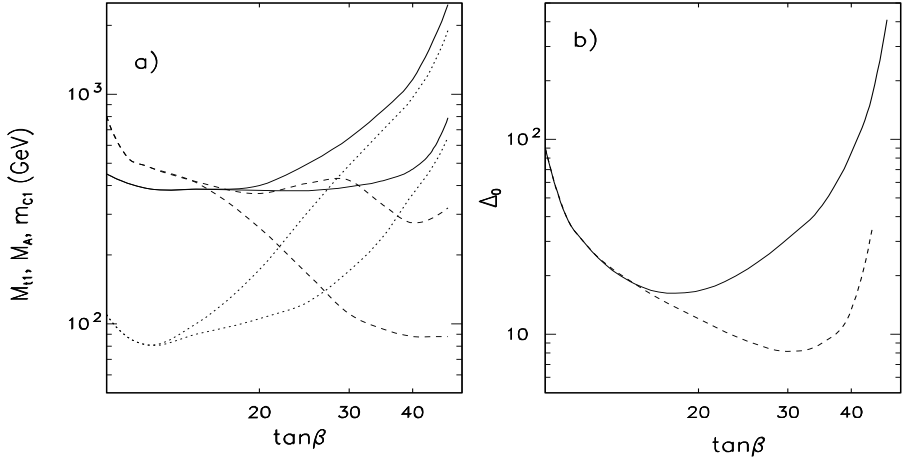,width=13.0cm,height=7.0cm}
\vspace{1.0truecm}
\caption{{\it a) Lower limits on the lighter (dotted lines) and heavier 
(solid lines) stop and on the $CP-$odd Higgs boson $A^0$ (dashed lines) 
in the minimal supergravity
scenario with $b-\tau$ Yukawa coupling unification, as 
functions of $\tan\beta$. Upper (lower) lines refer to the case with
the
$b\rightarrow s\gamma$ constraint imposed (not imposed).  b) The 
corresponding fine-tuning price with the $b\rightarrow s\gamma$ constraint 
imposed (solid line) and not imposed (dashed line).}}
\end{figure}

Returning now to the fine-tuning price, we show in Fig.~7b the dependence
of $\Delta_0$ on $\tan\beta$, with exact $b-\tau$ unification imposed
as an additional constraint on the parameter space. This dependence is 
shown both without and with $b\rightarrow s\gamma$ decay included among 
the experimental constraints, and we first focus on the case with 
$b\rightarrow s\gamma$ excluded. A comparison with Fig.~3 (where $b-\tau$ 
unification was not imposed) shows a substantial increase in the fine-tuning 
price for $\tan\beta=10$. This follows from the large values of $A_0$ 
needed in this case for $b-\tau$ unification (see the strong dependence of 
$\Delta_0$ on $A_0$ in Fig.~3) and from the simultaneous ease in satisfying 
the $b\rightarrow s\gamma$ constraint for $\tan\beta\approx10$ (as will be 
discussed shortly). On the other hand, for $\tan\beta=30$ it is easy to have 
$b-\tau$ unification. Hence, the price $\Delta_0$ to be paid without imposing 
the $b\rightarrow s\gamma$ constraint in Fig.~7b (dashed line) is essentially 
the same with or without $b-\tau$ unification (actually, it is slightly below 
the value seen in Fig.~4a for the case without $b-\tau$ unification but with 
the $b\rightarrow s\gamma$ constraint included).

We turn our attention now to a deeper understanding of the $b\rightarrow
s\gamma$ constraint and its interplay with $b-\tau$ unification. 
The first point we would like to make is that 
$b\rightarrow s \gamma$ decay is a rigid constraint in the minimal 
supergravity model, but is only an optional one for the general low-energy
effective MSSM. Its inclusion depends on the strong assumption that
the stop-chargino-strange quark mixing angle is the same as the CKM
element $V_{ts}$. This is the case only if squark mass matrices are diagonal
in the super-KM basis, which is realized, for instance, in the minimal
supergravity model. However, for the right-handed up-squark sector such an
assumption is not imposed upon us by FCNC processes \cite{MASI}.
Indeed, aligning the squark flavour basis with that of the quarks, 
the up-type squark right-handed  flavour off-diagonal mass squared
matrix elements
$(m^2_{\tilde U})^{13}_{RR}$ and $(m^2_{\tilde U})^{23}_{RR}$ 
are unconstrained by 
other FCNC processes. Therefore, in the limit that the other
flavour off-diagonal matrix elements are zero, and for sufficiently small
$(m^2_{\tilde U})^{23}_{RR}$, the 
couplings of  charginos to stops and  strange quark read
\begin{equation}
{\cal L}_{int} \supset - \bar{s}
\left(c_R^{ij} P_R + c_L^{ij} P_L\right) C^-_j
{\tilde t}_i
\end{equation}
with
\begin{eqnarray}
c_R^{1j} &\approx&{e\over s_W}Z^{1j\star}_+ V^\star_{ts}\sin\theta_{\tilde t}
-Z^{2j\star}_+\left(h_tV^\star_{ts}+h_c
{(m^2_{\tilde U})^{23}_{RR}\over M^2_{\tilde c_R} - M^2_{\tilde t_1}}\right)
\cos\theta_{\tilde t}\nonumber\\
c_R^{2j} &\approx&{e\over s_W}Z^{1j\star}_+ V^\star_{ts}\cos\theta_{\tilde t}
+Z^{2j\star}_+\left(h_tV^\star_{ts}+h_c
{(m^2_{\tilde U})^{23}_{RR}\over M^2_{\tilde c_R} - M^2_{\tilde t_2}}\right)
\sin\theta_{\tilde t}\nonumber\\
c_L^{1j} &=& -h_bV^\star_{ts}Z^{2j}_-\sin\theta_{\tilde t}\phantom{aaaaaa}
c_L^{2j} = -h_bV^\star_{ts}Z^{2j}_-\cos\theta_{\tilde t}
\end{eqnarray}
where $Z^{ij}_\pm$ are  matrices diagonalizing the chargino mass matrix
(defined in~\cite{ROSIEK}), the $h_{t,b,c}$ are  Yukawa couplings
and the $\tilde t_i$ are  stop mass eigenstates:
$\tilde t_R = -\sin\theta_{\tilde t}\tilde t_2
+\cos\theta_{\tilde t}\tilde t_1$. The factor $(\Delta^Rm^2)_{23}$ can be 
considered as a free parameter of the low-energy MSSM.  Indeed, there exist 
GUT models
\cite{RABY} that predict the mixing factor in the vertex 
$\bar s \tilde t_k C_i^-$  
to be considerably different from the CKM matrix element $V_{ts}$. 
We conclude that only a small departure from the minimal supergravity model 
is sufficient to relax the $b\rightarrow s\gamma$ constraint, and it is 
interesting to study separately its impact on the fine-tuning price.

In the minimal supergravity model the dominant contributions to
$b\rightarrow s\gamma$ decay come from the chargino-stop and charged 
Higgs-boson/top-quark loops. For intermediate and large $\tan\beta$, one
can estimate these using the formulae of~\cite{BAGIbsg} in the approximation 
of no mixing between the gaugino and higgsinos,
i.e., for $M_W\ll max(M_2, |\mu|)$. We get \cite{BOOLPO} 
\begin{eqnarray}
{\cal A}_W&\approx&{\cal A}_0^\gamma{3\over2}{m_t^2\over M_W^2}f^{(1)}
                   \left({m_t^2\over M_W^2}\right)\\
{\cal A}_{H^+}&\approx&{\cal A}_0^\gamma{1\over2}{m_t^2\over M_{H^+}^2}
                      f^{(2)}\left({m_t^2\over M_{H^+}^2}\right)\\
{\cal A}_{C}&\approx&-{\cal A}_0^\gamma\left\{\left({M_W\over M_2}\right)^2
                \left[\cos^2\theta_{\tilde t}
                 f^{(1)}\left({M_{\tilde t_2}^2\over M_2^2}\right)
                +\sin^2\theta_{\tilde t}
                 f^{(1)}\left({M_{\tilde t_1}^2\over M_2^2}\right)\right]
                \right.\nonumber\\
	    && -\left({m_t\over2\mu}\right)^2
                \left[\sin^2\theta_{\tilde t}
                 f^{(1)}\left({M_{\tilde t_2}^2\over\mu^2}\right)
                +\cos^2\theta_{\tilde t}
                 f^{(1)}\left({M_{\tilde t_1}^2\over\mu^2}\right)\right]
                \label{eqn:susy}\\
&&-\left.{\tan\beta\over2}{m_t\over\mu}
                {m_tA_t\over M^2_{\tilde t_2} - M_{\tilde t_1}^2}
                \left[f^{(3)}\left({M_{\tilde t_2}^2\over\mu^2}\right)
                - f^{(3)}\left({M_{\tilde t_1}^2\over\mu^2}\right)\right]
                \right\}\nonumber
\end{eqnarray}
where $\tilde t_1(\tilde t_2)$ denotes the lighter (heavier) stop,
\begin{equation}
\cos^2\theta_{\tilde t}={1\over2}\left(1 + \sqrt{1-a^2}\right), \phantom{aaaaa}
a\equiv{2m_tA_t\over M^2_{\tilde t_2} - M_{\tilde t_1}^2}, \phantom{aaaaa}
{\cal A}_0^\gamma\equiv G_F\sqrt{\alpha/(2\pi)^3}~V^\star_{ts}V_{tb}
\label{alwaysnumber}
\end{equation}
and the  functions $f^{(k)}(x)$ given in \cite{BAGIbsg} are negative.
The contribution ${\cal A}_{C}$ is effectively proportional to the stop mixing
parameter $A_t$, and the sign of  ${\cal A}_{C}$ relative to  ${\cal A}_W$
and ${\cal A}_{H^+}$ is negative for $A_t\mu<0$. 

We can discuss now the interplay of the $b-\tau$ unification and 
$b\rightarrow s\gamma$ constraints. The chargino-loop contribution
(\ref{eqn:susy}) has to be small or positive, since the 
Standard Model contribution and the charged Higgs-boson exchange (both 
negative) leave little room for additional constructive contributions.
Hence, one generically needs $A_t\mu<0$. Since
$\mu<0$ for $b-\tau$ unification, both constraints together require 
$A_t>0$. This is in line with our earlier results for the proper 
correction to the $b$ mass for $\tan\beta\simlt 10$, 
\footnote{This does not constrain the parameter space more than $b-\tau$
unification itself.  Note also that, if we do not insist on $b-\tau$
unification, the $b\rightarrow s \gamma$ constraint is easily satisfied since
$\mu>0$ is possible.} but typically in conflict with such corrections for 
larger values of $\tan\beta$. In the latter case, both constraints can be 
satified only at the expense of heavy squarks (to suppress a positive $A_t$ 
correction to the $b$-quark mass or a negative $A_t$ correction to 
$b\rightarrow s \gamma$) and a heavy pseudoscalar $A^0$.
Hence we have to pay a higher fine-tuning price, as seen in  Figs.~7a,b.

\section{Linear Relations between MSSM Parameters}

The minimal supergravity model with universal soft mass parameters 
discussed so far 
is based on the assumption that scalar mass parameters are not independent 
of each other
(and similarly for gaugino masses). This has obvious implications for the 
question of fine tuning, which can only be considered once the set of
initial parameters is specified. In particular, one could relax the 
universality assumption and study the question of fine tuning for each 
sfermion
flavour separately, with $\Delta_{\tilde m_i}$ being a measure of the 
fine tuning.. An increase (decrease) in the number of initial parameters 
is not directly correlated with increase or decrease of the necessary
fine tuning. For instance, suppose the parameters $a_i$ and $a_j$
with derivatives $\Delta_{a_i}$ and $\Delta_{a_j}$ are assumed
to be not independent but linearly related: $a_i=c_{ij}a_j$. In
this new scenario, the fine tuning is measured by $\Delta_{a_ia_j}=
\Delta_{a_i}+\Delta_{a_j}$. The relative magnitudes of $\Delta_{a_ia_j}$, 
$\Delta_{a_i}$ and $\Delta_{a_j}$ depend on the relative signs of 
$\Delta_{a_i}$ and $\Delta_{a_j}$, 
and vary from one region of parameters to another.
However, as observed in
\cite{CEP} and indicated in Section 2, the scalar sector has little impact
on the overall fine tuning, since the derivatives $\Delta_{\tilde m_i}$
are 
generically smaller than the other derivatives\footnote{The important
implications
of relaxing universality for the Higgs-boson mass parameters in the 
large-$\tan\beta$ scenario (see next Section) has a different origin: it
helps to permit
electroweak symmetry breaking in a larger
part of the parameter space.}. Therefore, 
scenarios with correlated or uncorrelated scalar masses have similar 
fine tuning.

\begin{figure}
\psfig{figure=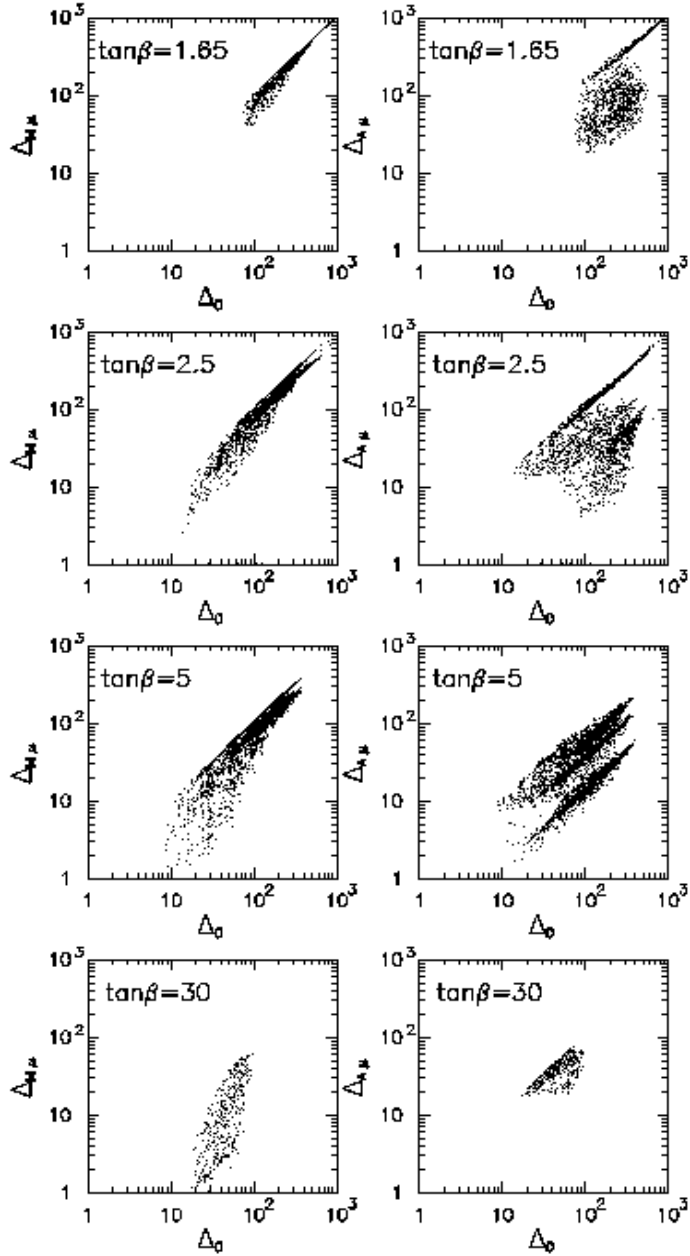,width=15.0cm,height=19.0cm}
\vspace{1.0truecm}
\caption{{\it Fine-tuning for correlated GUT-scale parameters 
($M_{1/2}=c_{M\mu}\mu_0$ in the left panels and 
$A_0=c_{A\mu}\mu_0$ in the right ones) versus fine tuning for uncorrelated
GUT-scale parameters for several values of  $\tan\beta$. Universal soft 
scalar masses are assumed at the GUT scale.}}
\end{figure}

\begin{figure}
\psfig{figure=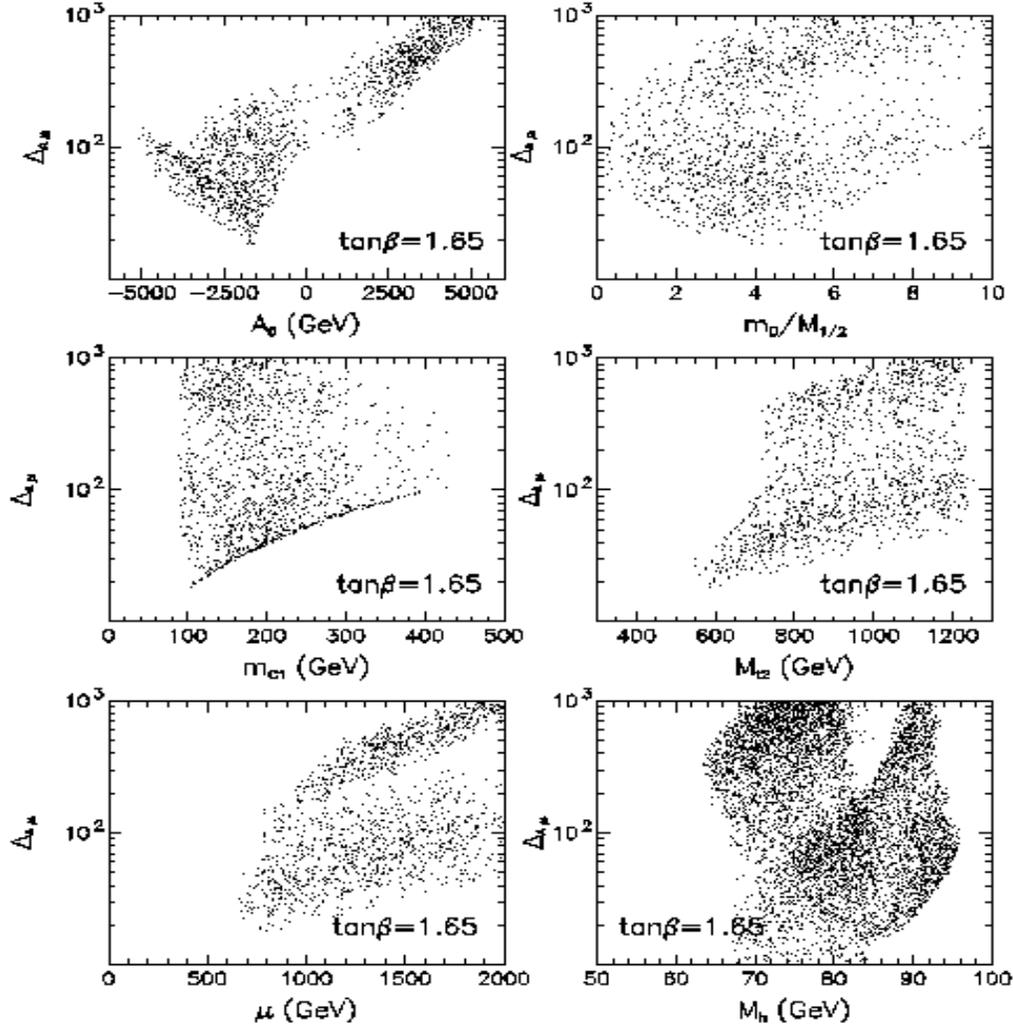,width=15.0cm,height=15.0cm}
\vspace{1.0truecm}
\caption{{\it As in Fig.~1, but assuming the linear correlation $A_0=c_{A\mu}\mu_0$ 
at the GUT scale.}}
\end{figure}

\begin{figure}
\psfig{figure=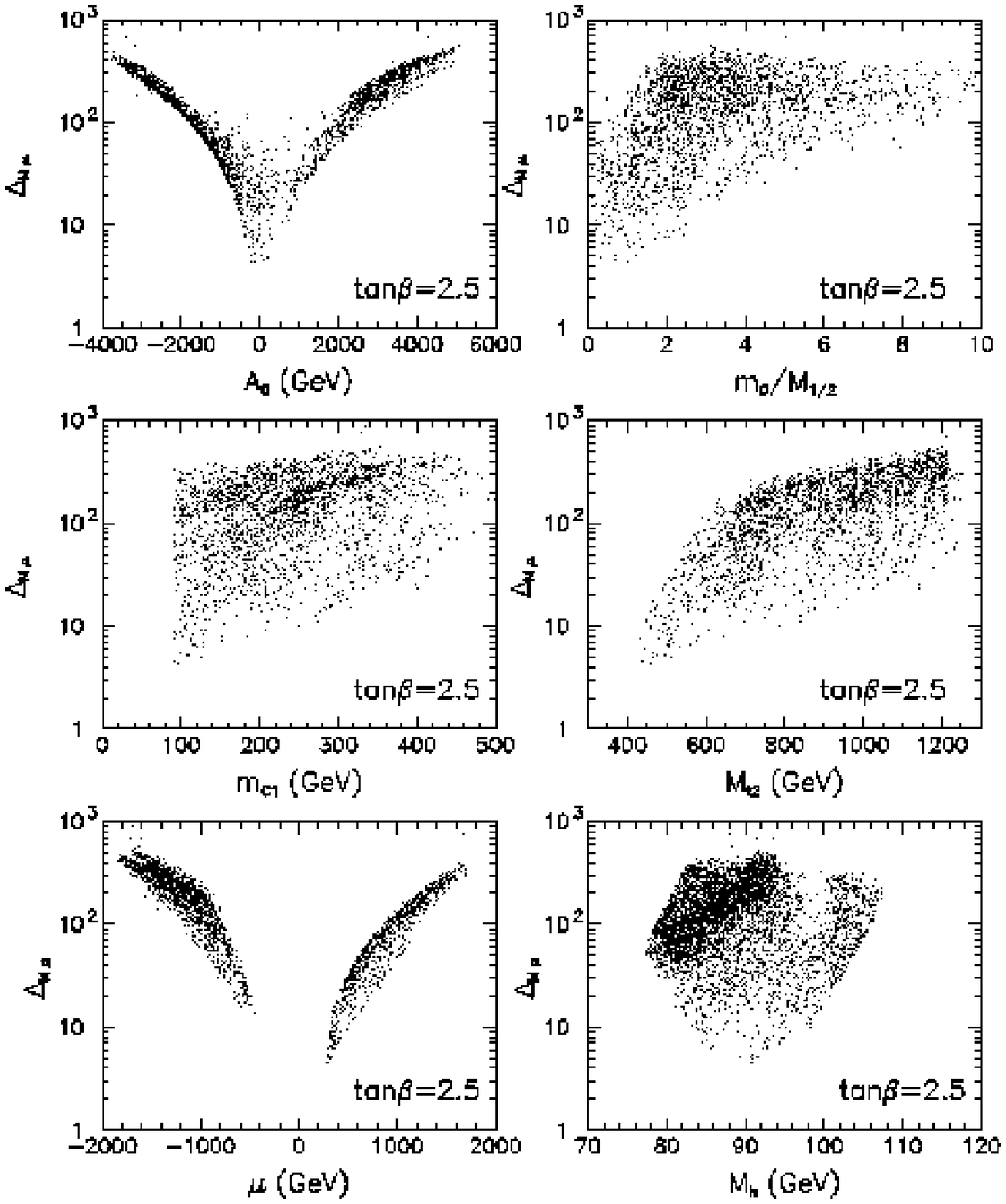,width=15.0cm,height=15.0cm}
\vspace{1.0truecm}
\caption{{\it As in Fig.~2, but assuming the linear correlation 
$M_{1/2}=c_{M\mu}\mu_0$ at the GUT scale.}}
\end{figure}

\begin{figure}
\psfig{figure=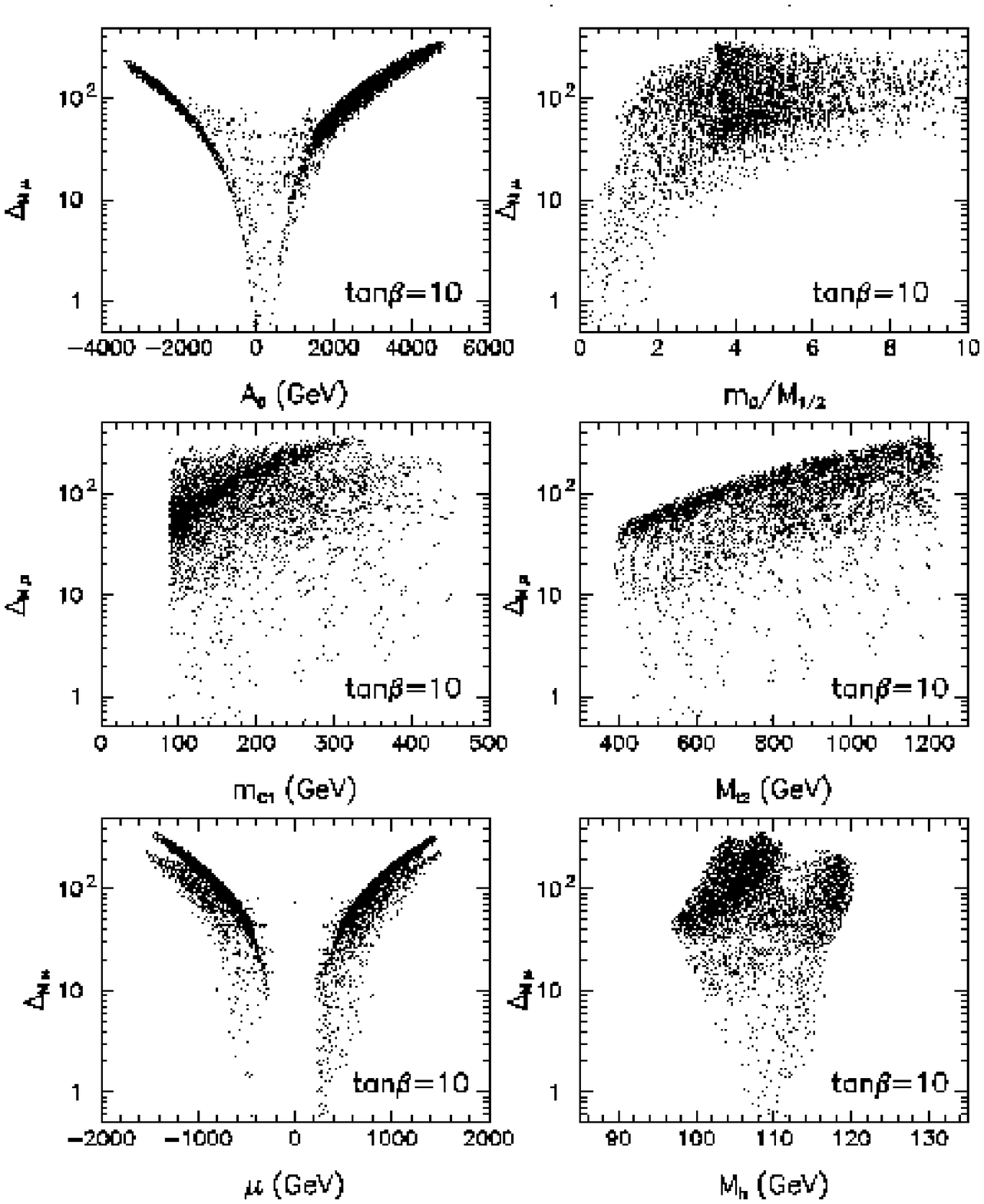,width=15.0cm,height=15.0cm}
\vspace{1.0truecm}
\caption{{\it As in Fig.~3, but assuming the linear correlation 
$M_{1/2}=c_{M\mu}\mu_0$ at the GUT scale.}}
\end{figure}

\begin{figure}
\psfig{figure=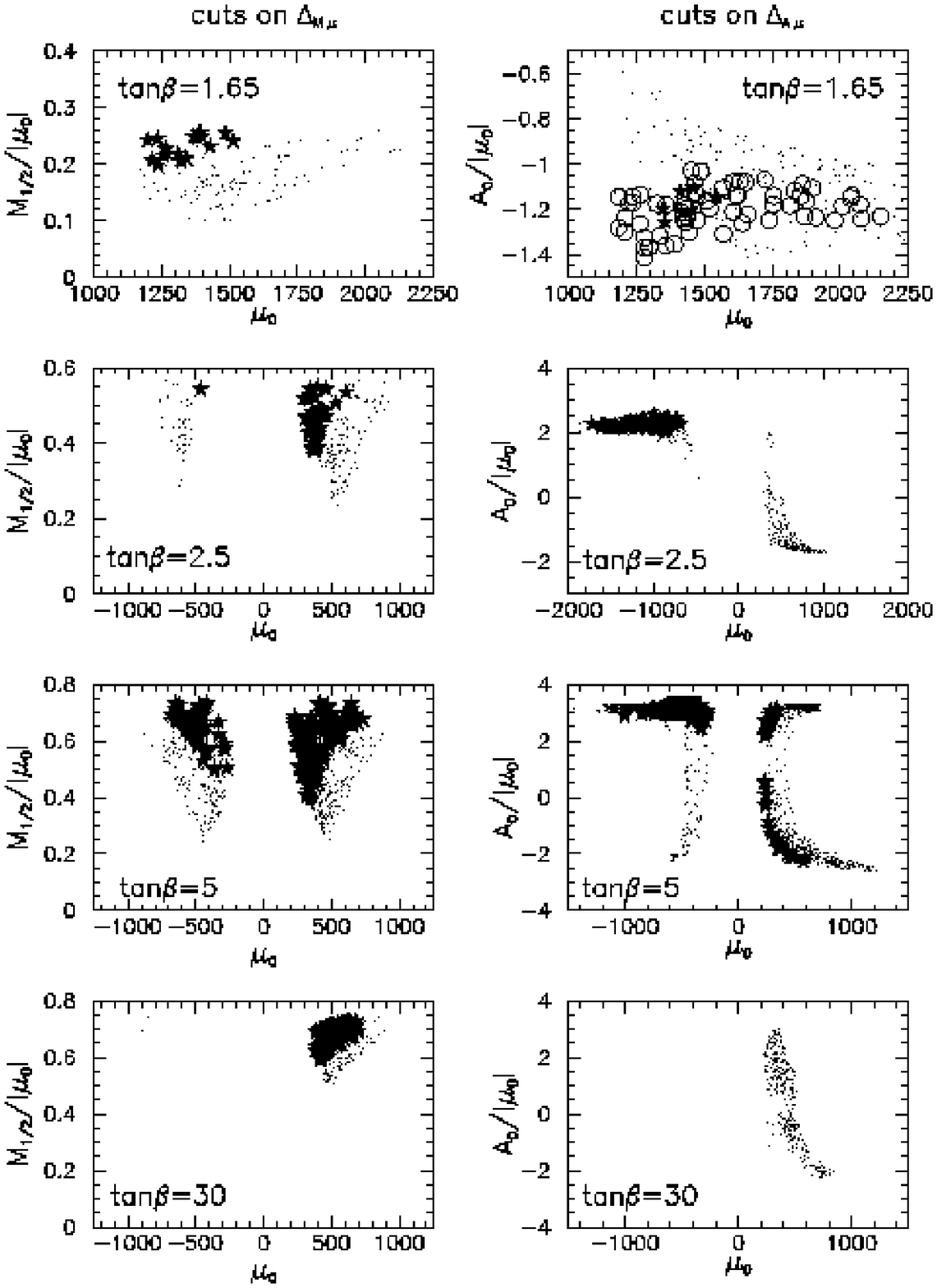,width=15.0cm,height=15.0cm}
\vspace{1.0truecm}
\caption{{\it Correlations between the GUT scale parameters for several
values of $\tan\beta$, assuming the following cuts on fine tuning.
For $\tan\beta=1.65$: in the left panel, 
parameter sets with $\Delta_{M\mu}<50$ are indicated by 
stars, and sets with $\Delta_{M\mu}<100$  by points, in the right panel,
sets with $\Delta_{A\mu}<20$ are indicated by by stars, $\Delta_{A\mu}<30$
by circles 
and $\Delta_{A\mu}<50$ by points. For $\tan\beta=2.5$ and 5: 
sets with $\Delta_{M\mu},\Delta_{A\mu} <10$ are indicated by stars and 
$\Delta_{M\mu},\Delta_{A\mu} <30$ by points. For $\tan\beta=30$: in the
left panel, sets with $\Delta_{M\mu}<3$ are indicated by stars and
$\Delta_{M\mu}<10$
by points, and in the 
right panel sets with $\Delta_{A\mu}<30$ are indicated by points.}}
\end{figure}

The discussion in Section 2 shows that most often the largest derivatives 
are $\Delta_{\mu_0}$, $\Delta_{M_{1/2}}$ and $\Delta_{A_0}$ and, moreover, 
in the phenomenologically relevant parameter space they are of opposite 
signs. Let us take as an example again small $\tan\beta$. 
We see in (\ref{eqn:univdeltas}) that $\Delta_{\mu_0}<0$, whereas 
$\Delta_{M_{1/2}}>0$ for $\mu>0$, which is
necessary to maximize $M_h$. Also, sign$(\Delta_{A_0})=-$sign$(\mu A_0)$,  
so that for $\mu>0$, sign$(\Delta_{\mu_0}\Delta_{A_0})$=sign$(A_0)$ and for 
negative $A_0$ the derivatives $\Delta_{\mu_0}$ and $\Delta_{A_0}$ are of 
opposite signs. Since negative $A_0$ is necessary for maximizing $M_h$, one 
expects that, by assuming there is some theoretical reason why some
of 
these parameters are not independent, one may significantly reduce 
the fine-tuning 
price. It is easy to study the simplest case of linear relations between 
these parameters. Linear relations are also enough to obtain substantial 
reductions in the fine tuning, as we find in our full 
one-loop numerical calculations.

\begin{figure}
\psfig{figure=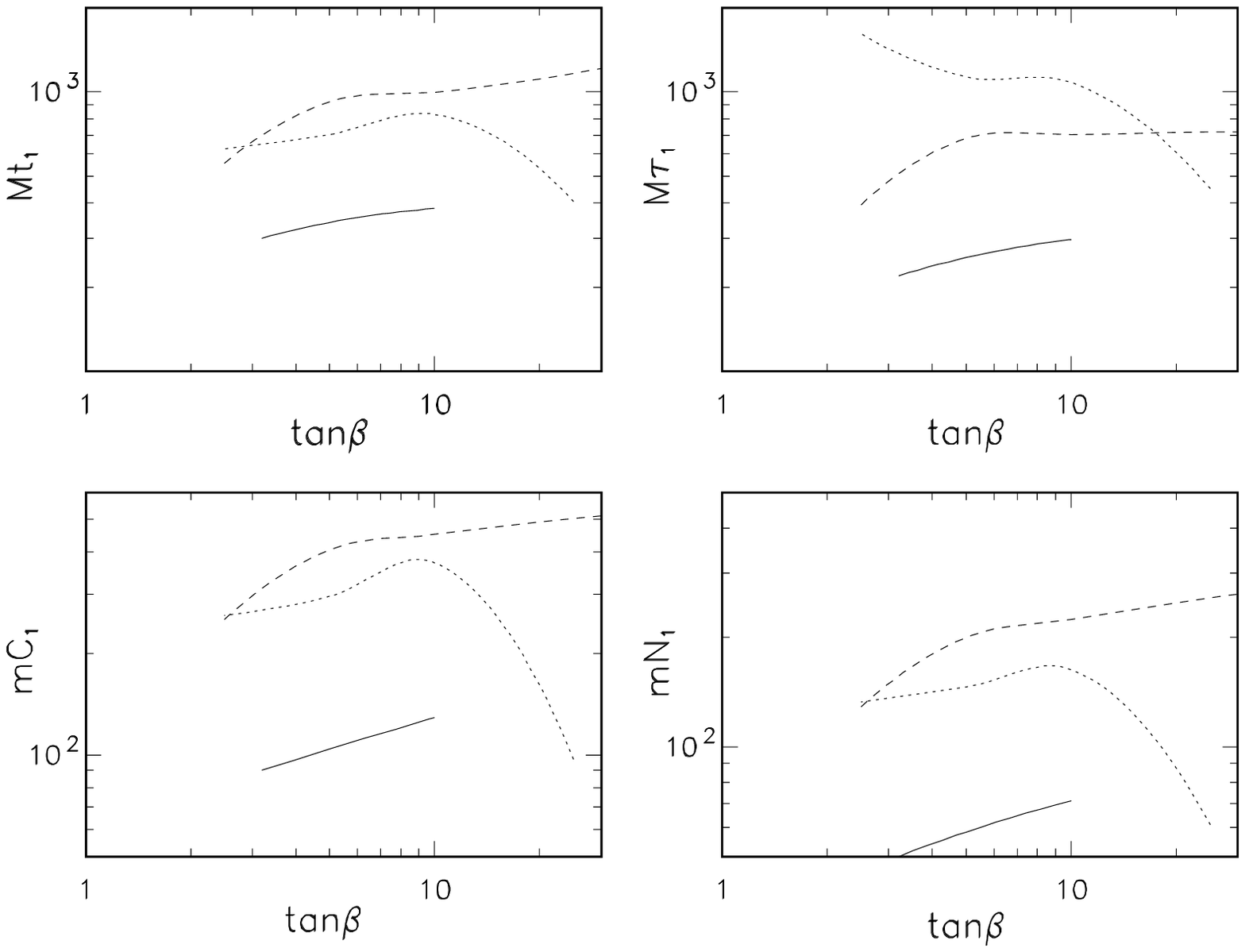,width=15.0cm,height=10.0cm}
\vspace{1.0truecm}
\caption{{\it Upper limit on the lighter stop (a), stau (b), chargino (c)
and 
neutralino (d) as functions of $\tan\beta$ for universal soft
scalar masses at the GUT scale requiring the fine-tuning 
measures $\Delta_0$ (solid),
$\Delta_{M\mu}$ (dashed) and $\Delta_{A\mu}$ (dotted) smaller than 10.}}
\end{figure}

We obtain the biggest reduction in the fine-tuning price by treating $M_{1/2}$ 
and $\mu_0$, or $A_0$ and $\mu_0$, as linearly related to each other, and the 
best choice depends on the value of $\tan\beta$. This is shown in Fig.~8, 
where we plot $\Delta_0$ versus $\Delta_{M_{1/2}\mu_0}$ and  
$\Delta_{A_0\mu_0}$ for several values of $\tan\beta$. For low $\tan\beta$ 
(close to the infrared quasi-fixed point) the best effect is obtained for 
the $A_0-\mu_0$ correlation, with the minimal 
fine tuning decreasing by a factor 
$\sim3.5$.
In Figs.~4c and 4e we plot minimal values of 
$\Delta_{M_{1/2}\mu}$ and $\Delta_{A_0\mu}$ as functions of $\tan\beta$
for
several assumed limits on $M_h$: $M_h>90$, 100, 105, 110 and 115 GeV, and
in Figs. 4d and 4f we show similar plots but for $M_h=95$, 100, 105, 110
and 115 GeV. The strong price reductions are evident. We see that 
$\Delta\sim{\cal O}(1)$ is
compatible with a large range of $\tan\beta$ values and Higgs boson masses.
In Figs.~9, 10 and 11, we plot $\Delta_{M_{1/2}\mu_0}$ 
and $\Delta_{A_0\mu_0}$ as 
functions of various mass parameters or physical masses, for several values of 
$\tan\beta$. The overall pattern remains similar to the $\Delta_0$ case,
but with an order of magnitude or more rescaling in the 
absolute values of the $\Delta$'s.
Nevertheless, putting some upper bound on the acceptable fine tuning remains
a strong constraint on the superparticle spectrum. This is more clearly
seen in Fig.~13, where we plot the upper bounds on several physical masses 
as a function of $\tan\beta$, as obtained by
requiring $\Delta_0<10$. In this plot we compare
the bounds obtained for $M_{1/2}-\mu_0$ and $A_0-\mu_0$ correlations with the 
bounds for the uncorrelated case ($\Delta_0<10$).  As is seen clearly
in Fig.~13, the upper bounds
are considerably relaxed if correlations are imposed, suggesting that it is
premature to use the fine-tuning price to derive
convincingly any upper mass limits, in the
absence of deeper theoretical understanding.

\section{String-Inspired Models}

So far, we have discussed linear dependences among soft
supersymmetry-breaking terms
in a  model-independent way. We now take a more theoretical
viewpoint, according to which the soft supersymmetry breaking parameters
are predicted by some physics at the GUT or string scale. 
It is likely that they will emerge from the high-scale theory
described in terms of more fundamental parameters. It is also 
plausible that the number of these parameters, at least of the relevant
ones, is smaller than the number of soft terms. The
latter will then not be independent. Such  scenarios indeed emerge in
various toy supergravity/string models for soft terms. 
The fine-tuning criterion would then require some revision:
even if the number of new parameters is not smaller than 
the number of soft terms discussed earlier, a reparametrization may
introduce more ``natural'' fundamental parameters.
Generically, in supergravity models, one can write the soft terms as
\beq
a_i = m_{3/2} f_i\left(p_\alpha\right)
\eeq
where $a_i$ is one of the soft 
supersymmetry-breaking terms ($M_{1/2}$, $m_0$, $A_0$, $B_0$),
or $\mu_0$, and the gravitino mass $m_{3/2}$ sets the overall mass scale. 
The functions $f_i(p_\alpha)$ are functions of dimensionless 
parameters $p_\alpha$ which can be regarded as, 
e.g., angles determining the
goldstino direction in the dilaton and moduli field space. 

Among the questions one can ask are: 
\begin{itemize}
\item[a)]
Does there exist a set of parameters $p_\alpha$ which is more natural than 
the soft terms themselves, and what are their properties?

\item[b)]
Are there any simple models with such parametrizations?
\end{itemize}
Within such a framework, we should study fine tuning with respect to the
parameters $m_{3/2}$ and $p_\alpha$. In the case of $m_{3/2}$,
simple dimensional analysis tells us that 
\beq
\Delta_{m_{3/2}} = 1
\eeq
On the other hand, the general formula for $p_\alpha$ is:
\bea
\Delta_{p_\alpha}={1\over(t^2-1)^2}
\sum_{ij}&&\!\!\!\!\!\!\!\left\{
-\left[(t^2+1){m_{3/2}^2\over M_Z^2} + 2t^2{m_{3/2}^2\over M_A^2}\right]
c_1^{ij}\right.\nn\\
&&\!\!\!-t^2\left[(t^2+1){m_{3/2}^2\over M_Z^2}+2{m_{3/2}^2\over M_A^2}\right]
c_2^{ij}\label{eq:Delta_p}\\
&&\!\!\!+\left.2t(t^2+1)\left[{m_{3/2}^2\over M_Z^2}+{m_{3/2}^2\over M_A^2}\right]
c_B^i c_\mu^j\right\}p_\alpha
\frac{\partial\left(f_i f_j\right)}{\partial p_\alpha}
\,.\nonumber
\eea
As an example, we now study a simple toy model~\cite{BRIBMUSC}
in which soft supersymmetry-breaking terms and the Higgs mixing 
parameter $\mu_0$ are described  at the GUT scale 
by the following parametrization:
\bea
M_{1/2}=&&\!\!\!\!\!\!\!\sqrt{3}m_{3/2}\sin\theta e^{i\gamma_S}\nn\\
A_0=&&\!\!\!\!\!\!\!-\sqrt{3}m_{3/2}\sin\theta e^{i\gamma_S}\nn\\
\left(m_{H_1}^2\right)_0 = &&\!\!\!\!\!\!\! \left(m_{H_2}^2\right)_0 =
m_{3/2}\left(1-3\cos^2\theta\left(\Theta_3^2+\Theta_6^2\right)\right)
\label{eq:BIMS}
\\
\mu_0=&&\!\!\!\!\!\!\!m_{3/2}\left(1+\sqrt{3}\cos\theta
\left(\Theta_3 e^{i\gamma_3} + \Theta_6 e^{i\gamma_6}\right)\right)\nn\\
B_0 \mu_0=&&\!\!\!\!\!\!\!2m_{3/2}\left(1 + \sqrt{3}\cos\theta
\left(\Theta_3\cos\gamma_3 + \Theta_6\cos\gamma_6\right)
+3 \cos^2\theta\cos\left(\gamma_3-\gamma_6\right)\Theta_3\Theta_6\right)\nn
\eea
where the angles $\theta$, $\Theta_i$ determine the goldstino direction 
in the dilaton/moduli parameter space, with $\sin\theta\approx1$ (0) 
corresponding to dilaton-- (moduli--) dominated supersymmetry breaking,
and the $\gamma_i$ are phases which we set to zero for simplicity in the 
rest of this discussion.

We now re-examine fine tuning in this new parametrization, considering 
first the sensitivity of $M_Z^2$ to $\sin\theta$. We obtain 
$\Delta_{\sin\theta}$ from (\ref{eq:Delta_p}) using the soft parameters 
(\ref{eq:BIMS}) and the coefficients $c^{ij}_k$ obtained by solving the 
one-loop renormalization-group equations. The resulting formula is very 
complicated (remember that $M_Z$, $M_A$ and $\tan\beta$ in (\ref{eq:Delta_p}) 
depend on the parameters $p_\alpha$) and will not be given here.  
It is, however, not very difficult to check that $\Delta_{\sin\theta}$
has typically several zeroes as a function of $\theta$ 
(for example, in the limit $y \to 1$ there are zeros for $\theta = n\pi/2$). 
Thus, there are regions in the new parameter space where the sensitivity 
to $\sin\theta$ is small. 
However, this does not mean yet that one can easily avoid fine tuning, 
since it is necessary to check whether the regions of small 
$\Delta_{\sin\theta}$ correspond to phenomenologically acceptable solutions. 
To do this, we analyse $M_Z^2$ itself as a function of $\theta$. 
Fig.~14 shows some typical plots of $M_Z^2(\theta)$. 

\begin{figure}
\psfig{figure=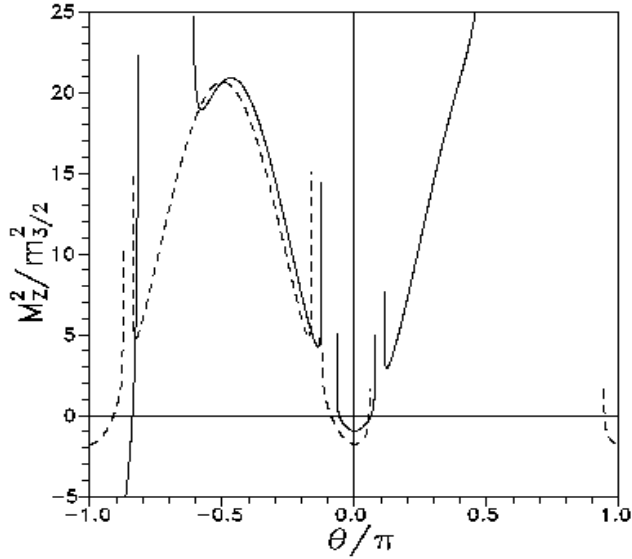,width=11.0cm,height=8.0cm}
\vspace{1.0truecm}
\caption{{\it Dependence of $M_Z$ on the angle $\theta$ in the string
parametrization of the soft terms shown in (20) for 
$\Theta_3=\Theta_6=-0.5$ (solid lines) and 
$\Theta_3=-\Theta_6=-0.5$ (dashed lines).}}
\end{figure}

We see that $M_Z^2(\theta)$ is either negative or rather large (in units 
of $m_{3/2}^2$) and positive at the extrema, and hence not acceptable. 
Experimental lower bounds on the masses of superpartners imply that 
the scale of supersymmetry breaking measured by $m_{3/2}$ must be rather 
big compared to the weak scale. From a phenomenological point 
of view,  the interesting regions of the parameter space are only those 
which give positive 
but rather small values of $M_Z^2$. Unfortunately, $\Delta_{\sin\theta}$ 
is never very small in such regions. 

We have checked this by a numerical calculation. We have scanned the 
($\theta$, $\Theta_3$, $\Theta_6$) parameter space looking for solutions with 
$M_Z^2$ between 0 and $m_{3/2}^2$ and with $M_A > 0.6 M_Z$. Such 
solutions exist only in quite a small part of the ($\Theta_3$, $\Theta_6$) 
parameter space. Moreover, they give very small values of $\tan\beta$ 
(quite close to 1), and have $\Delta_{\sin\theta}$ always well above 100. 
Thus, we conclude that the parametrization (\ref{eq:BIMS}) cannot solve the 
fine-tuning problem.

A search for more attractive models for soft terms, perhaps guided by the 
phenomenological discussion of Section 5, is certainly very important.

\section{Large-$\tan\beta$ Region}

In this Section we update the status of scenarios with (at least approximate) 
$t-b-\tau$  Yukawa coupling unification \cite{OLPOplb1,DIMRAB} and discuss 
their fine-tuning aspects. Such a possibility is realized, for instance, 
in $SO(10)-$type models. For $m_t=175$ GeV, $t-b-\tau$ unification 
predicts large values of $\tan\beta$: $\tan\beta\approx50$, and the Higgs 
boson mass $M_h\simgt 110$ GeV. Clearly, if the Higgs boson is not found 
at LEP~2, the phenomenological relevance of the
large $\tan\beta$ region will be accentuated.

Phenomenological properties of the large $\tan\beta$ region are well 
understood \cite{OLPO,HARASA,CAOLPOWA,DRENO,OLPOplb2}. 
Important aspects are 
the breaking of the electroweak symmetry and supersymmetric one-loop 
corrections to the bottom quark mass and to $b\rightarrow s\gamma$ decay.
To organize our discussion, let us begin with exact $t-b-\tau$ unification
of Yukawa couplings in the minimal supergravity model. The results of Section
5 can be readily used to conclude that it is not a realistic scenario
\cite{CAOLPOWA}. It is sufficient to observe that, in the parameter space
constrained by requiring proper electroweak symmetry breaking, it is impossible
to obtain sufficiently small one-loop supersymmetric corrections to the
$b$-quark mass for stop masses up to ${\cal O}(10$ TeV) 
(as can be estimated from (\ref{eqn:mbcorr}).
The problem is even worse if we try to be
consistent with $b\rightarrow s\gamma$ decay.

The question of some interest is how far we have to depart from  exact
unification of all three couplings in the minimal supergravity model to obtain
a more realistic parameter space. One way of answering this question is to 
impose $b-\tau$ unification, and to study the minimal values of the stop
masses and of the fine-tuning measure $\Delta_0$ which are necessary to
satisfy all the remaining constraints, as a function of $\tan\beta$.
This is shown in Fig.~7a and 7b, respectively, requiring the correct value
of the $b$-quark mass and, optionally, the correct $BR(b\rightarrow s\gamma)$.
As we discussed in Section 4, the prediction for the latter can be modified
by a departure from the minimal supergravity model that admits some flavour
structure in the stop mass matrices. We see from Fig.~7a that correct $M_b$
requires $M_{\tilde t_2}\sim 0.75(1.7)$ TeV for $\tan\beta$=45 without (with)
the $b\rightarrow s\gamma$ constraint included. The corresponding values
of $\Delta_0$ \footnote{For large $\tan\beta$ it is important to consider
the derivatives of $M_Z$ and $\tan\beta$, since the latter are proportional
to $\tan\beta$ and can be large.} are 170 and 35 respectively.
For $m_t=175$ GeV, $\tan\beta$=45 corresponds to $Y_t/Y_b\approx1.6$
(for $M_{\tilde t_i}\sim1$ TeV). 
The origin of all
these results is the extremely constraining role played by electroweak
symmetry breaking in the minimal supergravity model with $Y_t\approx Y_b$.
In the limit $Y_t=Y_b$, the two soft Higgs boson masses $m^2_{H_1}$ and 
$m^2_{H_2}$ run almost in parallel, and electroweak symmetry breaking
occurs
only for very large values of $M_{1/2}$ and $\mu$.

It has been pointed out in \cite{OLPOplb2} that a qualitatively new
situation appears in the large $\tan\beta$ scenario if we relax the 
universality of the Higgs-doublet soft mass parameters 
\cite{POPO,MANI,OLPOplb2}. This is because
the correlation $\mu\gg M_{1/2}$ is no longer necessary for proper 
electroweak symmetry breaking, and one obtains solutions with 
$\mu\sim M_{1/2}\sim{\cal O}(M_Z)$: as discussed in \cite{OLPOplb2}, 
the hierachy $m^2_{H_1}\gg m^2_{H_2}\simgt m^2_0$ is necessary for this. 
In consequence, 
in this scenario the supersymmetric loop corrections (\ref{eqn:mbcorr})
to $m_b(M_Z)$ can be small. It is, therefore, 
interesting to repeat the analysis in this case. Since it is easy to 
obtain acceptable physical $M_b$, even for $Y_t=Y_b=Y_\tau$, we restrict 
our analysis to this case. We study the case $\tan\beta=50$ and 
impose $M_b=4.8\pm0.2$ GeV, i.e., $2.72-3.16$ GeV for $m_b(M_Z)$.

\begin{figure}
\psfig{figure=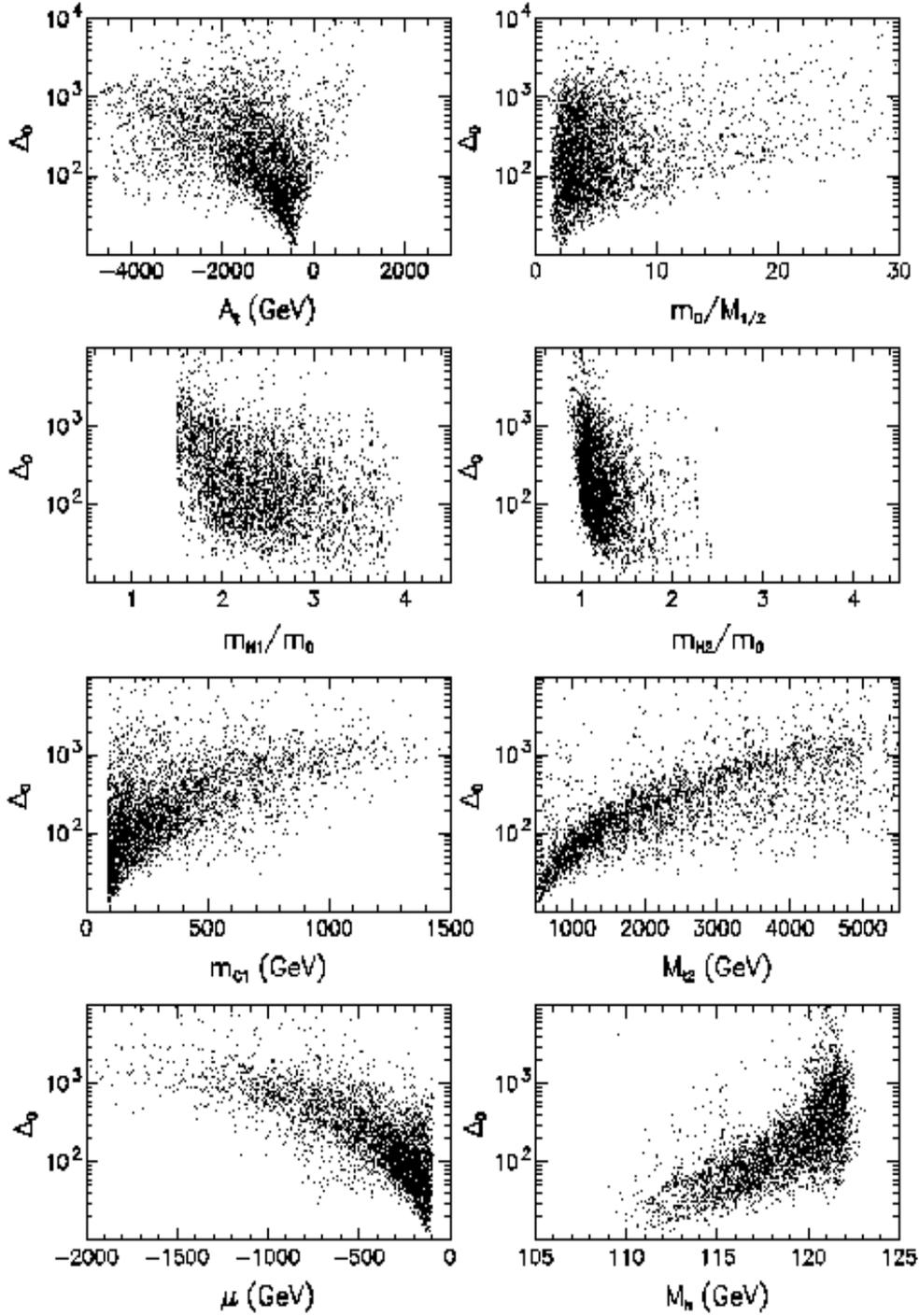,width=15.0cm,height=20.0cm}
\vspace{1.0truecm}
\caption{{\it The price of fine tuning for $\tan\beta=50$ and $t-b-\tau$
Yukawa coupling unification, as a function of various variables in models
with non-universal Higgs boson masses. The $b\rightarrow s \gamma$ 
constraint is 
not included.}}
\end{figure}

\begin{figure}
\psfig{figure=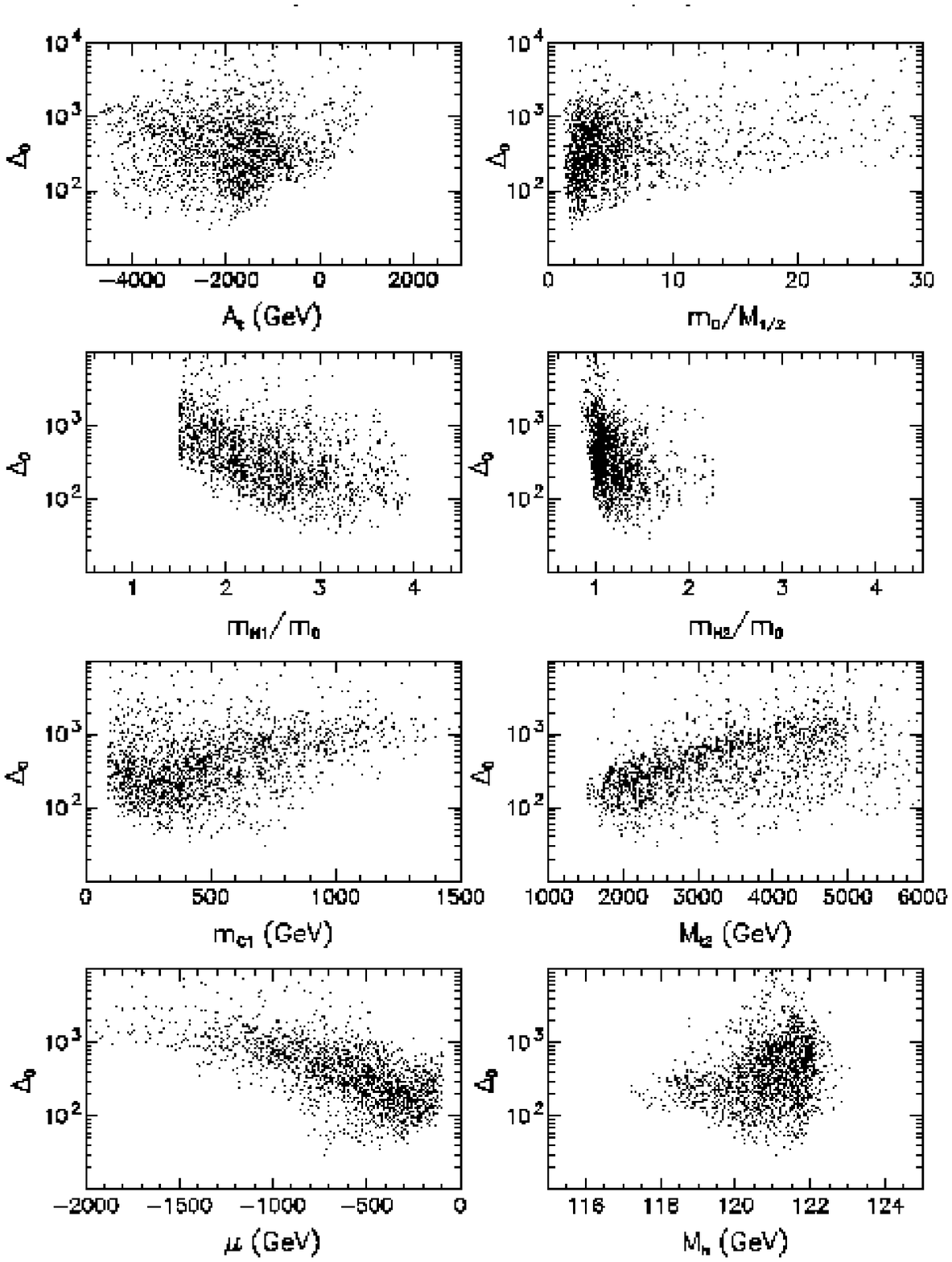,width=15.0cm,height=20.0cm}
\vspace{1.0truecm}
\caption{{\it The same as in Fig.~15, but with the $b\rightarrow s\gamma$
constraint included.}}
\end{figure}

In Fig.~15 we show some results for $\Delta_0$ as a function of several mass
parameters, with all the constraints included except for $b\rightarrow
s\gamma$. Only $\mu<0$ is possible since, as is clear from Fig.~5a, only
negative one-loop supersymmetric corrections are compatible with the correct
bottom-quark mass. Moreover, the correction has to be small enough and,
therefore, $A_t$ tends to be negative and
squarks must be relatively heavy. Due to the hierachy $m^2_{H_1}\gg m^2_0$,
the pseudoscalar $A^0$ is heavy enough to assure a small amount of fine
tuning~\footnote{Typically the dominant derivatives are
$(a_i / \tan\beta)(\partial\tan\beta / \partial a_i) \approx
-\tan\beta ((Bc^i_{\mu}a_i+\mu c^i_Ba_i) / M^2_A)$.}:
$\Delta_0\sim 10$ is possible. This result makes the 
large $\tan\beta$ region quite acceptable from the naturalness point of
view.

If we insist on being consistent with  $b\rightarrow s \gamma$ decay, 
the fine-tuning price increases to $\Delta\simgt 40$, as seen in Fig.~16.
This happens for reasons similar to those discussed for $\tan\beta = 30$ 
in the minimal model, and the discussion at the end of Section 4 applies 
unchanged to the present case.

Finally, we remark that the non-universal Higgs boson masses discussed here
give solutions with higgsino-like neutralinos. Thus, such scenarios generically
lead to a rather low neutralino dark matter density~\cite{ELFAGAOLSC}.

\section{Conclusions}

Comparing the situation before and after LEP, the fine-tuning price in
the minimal supergravity model has increased significantly, largely
as a result of the unsuccessful Higgs boson search. Comparing
different values of $\tan\beta$, we find that naturalness favours an
intermediate range. Fine tuning increases for small values because
of the lower limit on the Higgs mass, in particular, and increases
for large values because of the difficulty in assuring correct
electroweak symmetry breaking.

Additional theoretical assumptions may have a significant impact on
the fine-tuning price. For example, requiring
$b-\tau$ Yukawa-coupling unification would increase the price
significantly at intermediate $\tan\beta$, whereas imposing certain
linear correlations between mass parameters could diminish it
substantially. One particular class of models imposing such
have been motivated from string constructions: unfortunately,
those currently available
do not seem to reduce the fine-tuning price significantly,
so naturalness considerations do not favour these models to any
substantial extent. However, the
search for realistic theoretical models which do reduce the
fine-tuning price is a very interesting issue.

We have found that $b\rightarrow s\gamma$ decay is a potentially
important constraint for large $\tan\beta$, but we would argue
that it should be regarded as optional. The flavour structure
of squark couplings could differ from those of the quarks, and there are 
no direct FCNC limits on flavour violation among superpartners of the
up quarks.

A final comment concerns the region of very large $\tan\beta$.
In this case, the fine-tuning price can be reduced quite
substantially by allowing non-universal soft mass
parameters for the Higgs bosons. This is in contrast to the situation
at lower $\tan\beta$, where non-universal mass parameters do not reduce
the price significantly.

We re-emphasize that naturalness is subjective criterion, based on
physical intuition rather than mathematical rigour. Nevertheless,
it may serve as an important guideline that offers some discrimination
between different theoretical models and assumptions. As such, it
may indicate which domains of parameter space are to be preferred.
However, one should be very careful in using it to set any
absolute upper bounds on the spectrum. We think it safer to
use relative naturalness to compare different scenarios, as we have done
in this paper.

\vskip 0.3cm
\newpage
\noindent {\bf Acknowledgments.} 

\noindent The work of P.H.Ch. has been partly supported by the Polish State 
Committee for Scientific Research grant 2 P03B 030 14 (1998-1999) and by the 
U.S.-Polish Maria Sk\l odowska-Curie Joint Fund II (MEN/DOE-96-264). 
The work of M.O. was supported by the Polish State Committee for Scientific 
Research grant 2 P03B 030 14 (1998-1999) and by the Sonderforschungsbereich 
375-95 f\"ur Astro-Teilchenphysik der Deutschen Forschungsgemeinschaft.
The work of S.P. has been supported by the Polish State Committee for 
Scientific Research grant 2 P03B 030 14 (1998-1999).

\end{document}